\documentclass[superscriptaddress,aps, reprint,
 amsmath,amssymb,
 aps,
]{revtex4-2}

\usepackage{graphicx}
\usepackage{dcolumn}
\usepackage{bm}

\usepackage{float}
\usepackage{soul}
\usepackage{amsmath}
\usepackage{braket}

\begin{document}


\title{Quartz phononic crystal resonators for hybrid acoustic quantum memories}
\author{Yang Hu}
\affiliation{Department of Physics and Astronomy, University of Pittsburgh, Pittsburgh, PA, USA}
\author{Angad Gupta}
\affiliation{Department of Physics and Astronomy, University of Pittsburgh, Pittsburgh, PA, USA}
\author{Jacob Repicky}
\affiliation{Department of Applied Physics, Yale University, New Haven, CT 06511, USA}
\author{Michael Hatridge}
\affiliation{Department of Applied Physics, Yale University, New Haven, CT 06511, USA}
\author{Thomas P. Purdy}
\email[Contact Author: ]{tpp9@pitt.edu}
\affiliation{Department of Physics and Astronomy, University of Pittsburgh, Pittsburgh, PA, USA}

\date{\today}
\begin{abstract}
Circuit quantum acoustodynamics systems have emerged as a promising platform for quantum information by coupling superconducting qubits to mechanical resonators, with their long-lived mechanical modes serving as quantum memories. We demonstrate suspended quartz phononic crystal resonators at 100~MHz with millisecond lifetimes at 8~K. With a contactless electrode geometry suppressing both two-level system losses and other electrode-induced energy dissipation, we evaluate the piezoelectric coupling rate between the mechanical modes and fluxonium qubits (resonant coupling) and transmon qubits (parametric coupling mediated by a Josephon-junction-based three-wave mixer). We further discuss multi-period defect geometries for enhancing these coupling rates.
\end{abstract}

\maketitle


\section{\label{sec:level1}Introduction}
Processing quantum information for computation and networking protocols requires strongly interacting and coupled systems that are generally susceptible to loss and decoherence, posing significant challenges to scaling quantum architectures. To address this problem, hybrid quantum memories, which are units that temporarily store quantum information for later on-demand retrieval, have emerged as a crucial building block for quantum information processors \cite{ Superconducting_circuit_QuantumInformation_outlook_Devoret_Schoelkopf_2013, Quantum_RAM_2019}, networks \cite{The_quantum_internet_Kimble_2008, Quantum_internet_review_2018}, and metrology and sensing \cite{QuantumMem_for_quantumSensing_Zaiser_Wrachtrup_2016, QuantumMem_review_Simon_Müller_et_al._2010}.  
Diverse quantum memory platforms are under investigation including solid state spins \cite{QuantumMem_for_quantumSensing_Zaiser_Wrachtrup_2016, QuantumMem_rare_earth_ion_Thiel_Cone_2011}, atomic ensembles \cite{Entangle_of_two_quantumMem_atomEnsemble_Yu_JianWeiPan_2020}, 3-D cavities \cite{Reagor_2016, QuantumMemory_3D_cavity_Brecht_2016,Compact_3D_quantum_memory_2018}, on-chip microwave resonators \cite{Quantum_memory_on_chip_resonator_Naik_2017,Quantum_memory_on_chip_resonator_Sardashti_2020}, and mechanical resonators \cite{Qubit_storage_with_resonator_Cleland_2004}. Among these candidates, mechanical resonators are attracting increasing interest due to their long lifetimes and compactness \cite{Quantum_memory_A.N.Clealand_2010, Quantum_AcousticsWithSC_qubit_YiwenChu_Schoelkopf_2017,Ultralong_phonon_lifetime_2020} enabled by the slow acoustic speed in solids compared to the photon speed, making them advantageous for building scalable quantum systems. Hence, circuit quantum acoustodynamic systems (circuit-QAD), which couple superconducting qubits to mechanical resonators, have emerged as a promising platform for manipulating mechanical states and storing and processing quantum information \cite{Quantum_RAM_2019, QC_catCode_Chamberland_2022, Pechal_2019, Linear_mechanical_QC_Qiao_2023,Yang_Kladarić_Drimmer_Lüpke_Lenterman_Bus_Marti_Fadel_Chu_2024}.

In this work, we experimentally demonstrate millisecond-lifetime quartz suspended 1-D phononic crystal (PnC) resonators at 100~MHz frequency that are suited for piezoelectric coupling to superconducting circuits. Phononic crystal resonators (PCRs) \cite{PnC_review_2021} consist of suspended periodic structures that act as acoustic Bragg mirrors to confine mechanical modes in a central defect unit cell (see Fig.~\ref{PnC_figure}).  This acoustic shielding suppresses energy radiation into the bulk at the anchor points, allowing PCRs to reach their intrinsic dissipation limit~\cite{Ultralong_phonon_lifetime_2020, schmid_fundamentals_2023, gokhale_approaching_2017}. Other geometries of mechanical devices, such as bulk acoustic wave resonators ~\cite{Goryachev_Tobar_2012,Kharel_Rakich_2018,Quantum_memory_A.N.Clealand_2010,Quantum_AcousticsWithSC_qubit_YiwenChu_Schoelkopf_2017} and surface acoustic wave (SAW) resonators \cite{gustafsson_2014, manenti2017,satzinger2018, SAW_mechanical_resonator_Emser_Lehnert_2022,Linear_mechanical_QC_Qiao_2023} are typically limited by scattering into the bulk from surface features and roughness and generally have a much larger footprint.  One-dimensional PCRs have been fabricated from a variety of piezoelectric materials, including LiNbO$_3$~\cite{arrangoiz2019,Strong_dispersive_coupling_with_fluxonium_2023}, AlN~\cite{Vainsencher2016}, quartz~\cite{emser_thin-film_2024}, GaP~\cite{stockill2022}, and GaAs~\cite{balram2016}, and some devices use a hybrid approach with piezoelectric transducers or electrodes off the device~\cite{balram2016,Arrangoiz_Arriola2018,mirhosseini2020,weaver2024}.  We select quartz as the material for our PCRs because of its low intrinsic loss \cite{CRA_impurity_Zimmer_Tünnermann_2007,Galliou_Abbé_2011,Goryachev_Tobar_2012}, which enables acoustic resonators with high quality factor (Q)  \cite{Galliou_Tobar_2013, Kharel_Rakich_2018}. While quartz BAW \cite{Galliou_Tobar_2013, Kharel_Rakich_2018} and SAW resonators \cite{Sletten2019,SAW_mechanical_resonator_Emser_Lehnert_2022} have been actively studied in the context of circuit-QAD, quartz PCRs have only recently begun to be explored in Ref.~\cite{emser_thin-film_2024}, partially due to the challenges associated with accessing high-quality thin-film crystalline quartz \cite{Carreter_Genevrier2013,Quartz_resonators_Sohn_Lončar_2017}, which motivates this work.

Rather than working directly at the gigahertz frequencies of superconducting qubits, we gear down our resonator frequency $\omega_m$ to be around 100~MHz. This lower frequency optimizes the thermal decoherence for a given quality factor, $Q_m$. At high frequency $\hbar \omega_m\gg k_BT$, the thermal decoherence time $\tau_{th}\approx1/\Gamma_m=\frac{Q_m}{\omega_m}$ improves with decreasing frequency. But in the low frequency regime $\tau_{th}\approx 1/(\bar{n}_{th}\Gamma_m)\approx\frac{\hbar Q_m}{k_BT}$, which is independent of $\omega_m$ (here $\bar{n}_{th},\Gamma_m$ are average number of thermal phonon, mechanical damping rate and $T$ is the temperature) \cite{Cavity_optomechanics_2014}. At the 10~mK scale temperatures of a dilution refrigerator, this transition happens at $\sim$100s~MHz.  Additionally, lower frequencies allow larger (yet still microscopic) device dimensions, which reduce the surface-to-volume ratio, lowering the sensitivity to surface two-level systems (TLSs). The $\sim$10~$\mu$m device size scale enables us to stand off our electrodes on a separate dedicated chip hosting the superconducting circuits, rather than placing electrodes directly on the mechanical resonators \cite{Goryachev_Tobar_2012,Quantum_AcousticsWithSC_qubit_YiwenChu_Schoelkopf_2017,Arrangoiz_Arriola2018,Noncontact_coupling_2024}. This contactless-electrode approach eliminates mechanical dissipation in the electrodes and suppresses decoherence from TLSs introduced by electrodes and interfaces \cite{Q_of_Qz_resonators_near_4K_Galliou_Bourquin_2015, Loss_channel_Amir_2021, emser_thin-film_2024} while preserving coupling strength with minimal degradation \cite{Noncontact_coupling_2024}. With such contactless electrode architecture, we estimate the coupling strength of the mechanical mode to a fluxonium qubit with resonant coupling rate $g/(2\pi)\sim 100~\textrm{kHz}$ and to a transmon qubit with parametric coupling rate $g_{eff}/(2\pi)\sim 20~\textrm{kHz}$ (swap time $\sim 20~\mu s$) mediated by a Superconducting Nonlinear Asymmetric Inductive eLement (SNAIL) \cite{3-wave_mixing_SNAIL_Devoret_2017}. We further discuss methods to enhance the coupling rates to bring memory read and write operations to the microsecond time scale.

\section{\label{sec:level1} Device design and fabrication} 

\begin{figure}[h]
    \includegraphics[width = \linewidth]{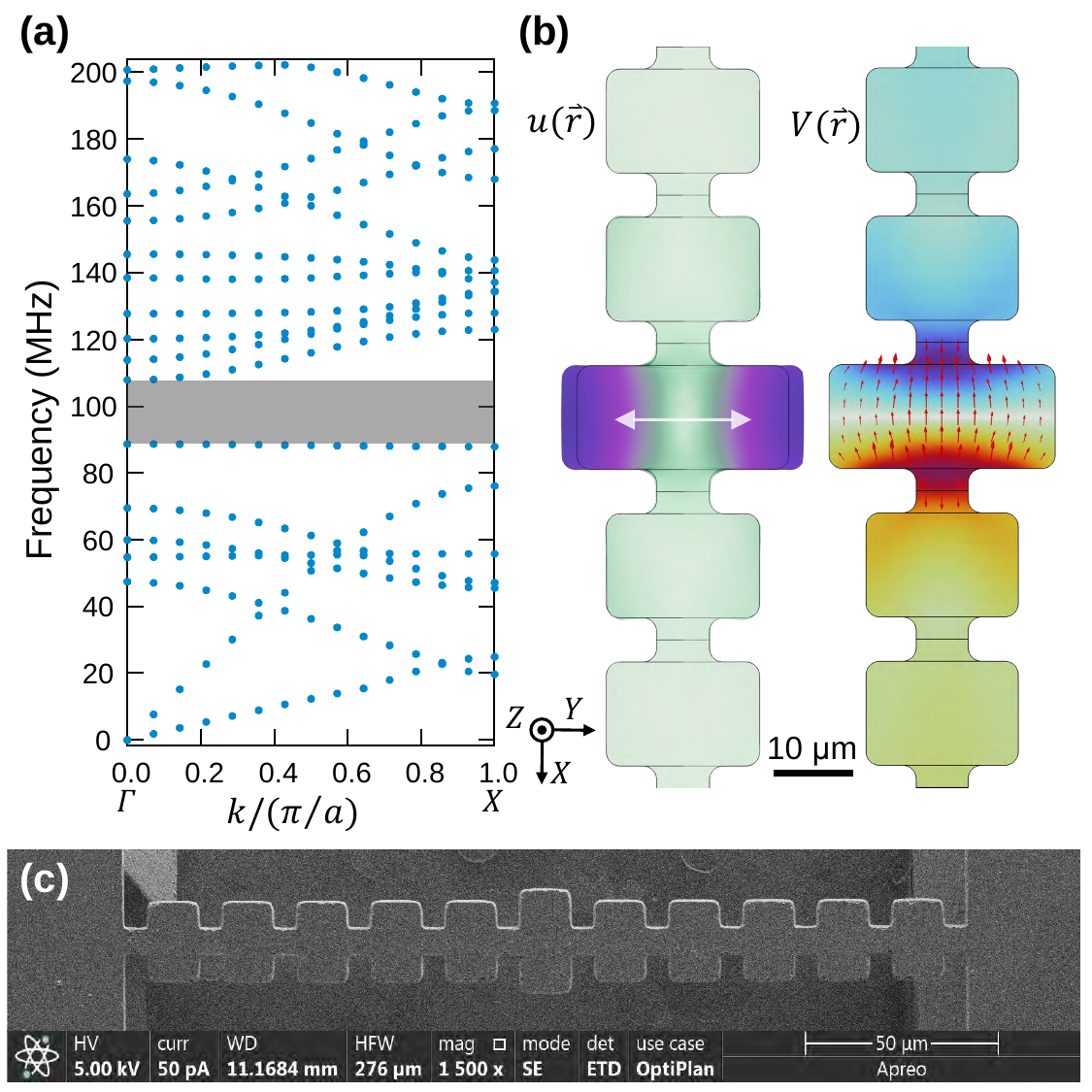}
    \caption{\label{Device} \textbf{Device simulation and fabrication.} (a) Simulated acoustic band structure of the phononic crystal Bragg mirror unit cell with lattice constant $a$. Gray shaded area indicates the $\sim$20~MHz complete acoustic band gap. (b) Finite-element simulation (COMSOL) of the width-extension mode showing the mechanical displacement (left) and electric potential (right) of the PnC resonator. The bottom left coordinate system depicts the crystal axes of the quartz. The white arrow indicates the direction of  motion, and the red arrows represent the electric field distribution. (c) SEM image (top view) of a fabricated PnC resonator.}
    \label{PnC_figure}
\end{figure}

Our devices are approximately 3.5~$\mu$m-thick, double-clamped, suspended, 1-D quartz PnC resonators fabricated from Z-cut $\alpha$-quartz bonded on a high-resistivity silicon substrate. The resonator is patterned along the crystal X-axis with both ends anchored to large rectangular pads supported by the underlying silicon (Fig.~\ref{Device}). Our periodic PnC Bragg mirrors are designed to exhibit a $\sim$20~MHz complete acoustic band gap centered around $\sim$100~MHz. A central rectangular block wider than the Bragg mirror cell is introduced as a defect. This defect creates a localized width-extension mode with strain along the crystal Y-axis, as shown in Fig.~\ref{Device}(b), along with the corresponding electric potential distribution. This mode permits strong piezoelectric coupling with electrode pairs positioned around the central defect along the crystal X-axis.

The device fabrication begins by thinning the quartz layer of a bonded quartz-on-silicon wafer from $\sim$70~$\mu$m down to target thickness ($\sim$3.5~$\mu$m),  using ICP-RIE (Inductively Coupled Plasma - Reactive Ion Etching) with SF\textsubscript{6} and Ar gases. Device patterns were then defined by photolithography and protected by a Ni hard mask. Subsequent RIE defines the resonator profile.  The RIE was continued to etch about 10~$\mu$m into the silicon substrate to facilitate device suspension.  Following removal of the Ni mask, KOH wet etching undercuts the Si and suspends the devices. Additional fabrication details are presented in Appendix~\ref{Appendix:Device_fabrication}.

We systematically investigate PnC resonator performance by fabricating devices with varying defect width (defect dimension transverse to the chain direction) and number of mirror cells. The mode frequency varies inversely with the defect width, allowing us access to modes both within and outside the phononic band gap.

\section{Experimental setup}

\begin{figure}
    \centering
    \includegraphics[width = \linewidth]{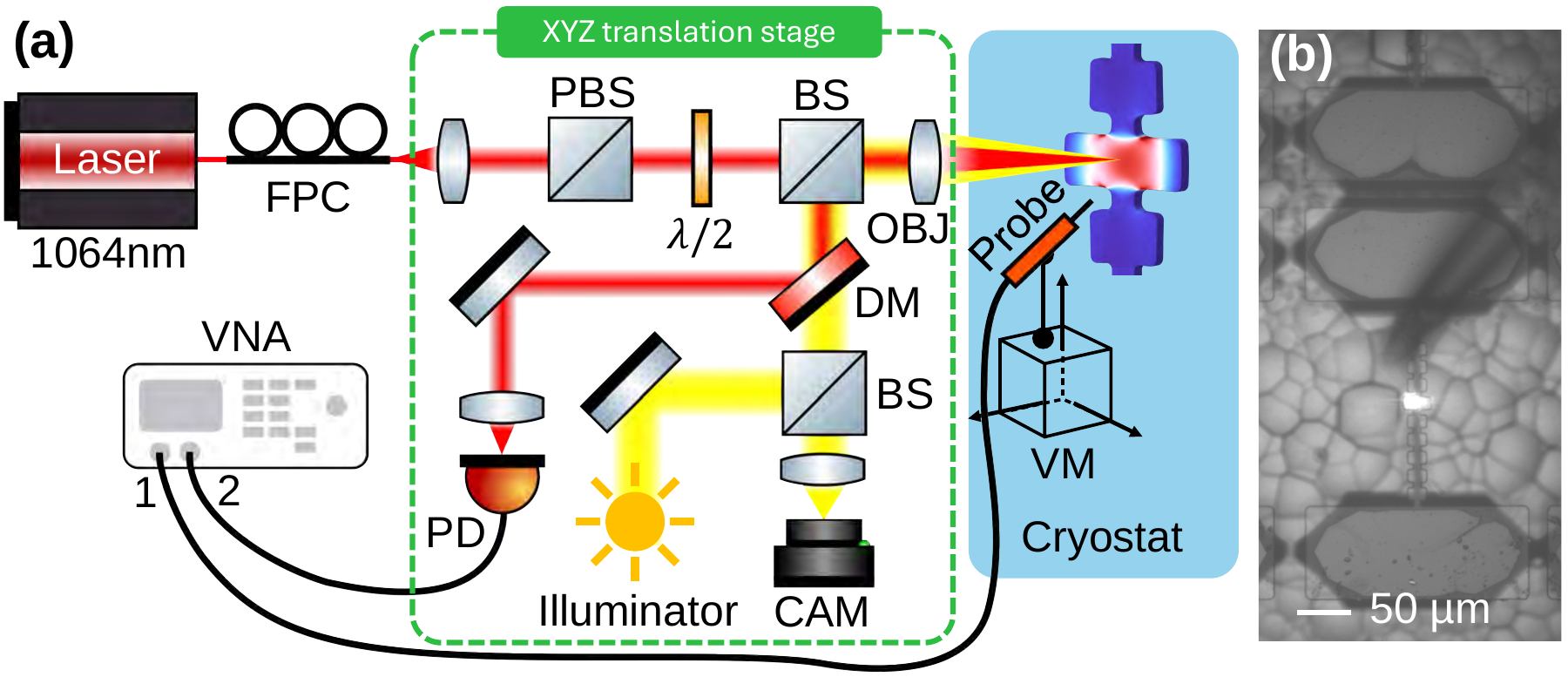}
    \caption{\textbf{Experiment setup}. (a) Schematic of the experiment setup. The setup includes a 1064~nm fiber-coupled laser, fiber polarization controller (FPC), polarizing beam splitter (PBS), beam splitter (BS), dichroic mirror (DM), half-wave plate ($\frac{\lambda}{2}$), vacuum manipulator (VM), Vector Network Analyzer (VNA), objective lens (OBJ), fiber optic illuminator, and camera (CAM). Blue-shaded region represents the chamber of cryostat. The simulated distribution of the dominant strain component ($S_{yy}$) of the width-extension mode is shown, with maximum strain localized at the central defect. (b) An optical microscope image of the tungsten tip positioned near the resonator and laser spot focused on the defect.}
    \label{Exp setup}
\end{figure}

We characterize our devices with a custom-built cryogenic probe station integrating both microwave and optical probes, utilizing the piezoelectric and photoelastic effects in quartz (see Fig.~\ref{Exp setup}). The position of both the optical probe and microwave probe can be independently scanned, with positioning facilitated by a microscope coaxial with the optical probe.

 The microwave probe consists of a room temperature tungsten tip (GGB T-4-35) soldered onto the center conductor of a coaxial cable, which is mounted on a three-axis vacuum manipulator. The probe tip is positioned near the mechanical resonator for mode excitation and is connected to port 1 of a Vector Network Analyzer (VNA). Although the scanning probe is highly undercoupled to our devices (as opposed to the planar electrode geometry described below), when paired with high-sensitivity optical detection, our system provides a flexible tool for rapidly characterizing large numbers of devices.  This testing setup does not require us to engineer dedicated, optimized electrodes suited for microwave reflection measurements, employ frequency multiplexing, or de-embed the influence of electrodes on the mechanics.

Simultaneously, a 1064~nm laser optical probe is focused onto the resonator (incident power around 4~mW) through a vacuum viewport and window on the cryostat radiation shield. A photodetector collects the reflected signal and sends its output to port 2 of the VNA. Due to the photoelastic effect, the strain of the mechanical modes, driven by the microwave probe, modulates the quartz refractive index. Such modulation induces a beating signal at the mechanical mode frequency in the reflected optical power. Due to the polarization dependence of the photoelastic effect, we control the input laser polarization using a fiber polarization controller, a free space polarizing beamsplitter, and a half-wave plate, before focusing the beam onto the device with a 10$\times$ objective lens (Mitutoyo M Plan Apo NIR 10X, NA = 0.26, spot size $\approx$ 3~$\mu$m). For our Z-cut quartz, the detected signal is maximized when the optical polarization is orthogonal to the mechanical strain (see Appendix~\ref{Appendix:Photoelastic_characterization} for more details). The entire free-space optical probe and microscope setup is mounted on a three-axis translation stage for manual positioning.

Our devices are mounted inside a helium-flow cryostat (Janis ST-100). The resonator chip is fixed onto a sapphire spacer with N-grease, which is clamped onto a copper sample holder mounted on the cold finger of the cryostat. The sapphire spacer is introduced to isolate the resonator from the ground plane of the copper sample holder. A silicon diode temperature sensor is attached to the sample holder. Blackbody radiation from the microwave probe setup prevents the devices from reaching the base temperature of the cryostat. To address this problem, a 50~$\mu$m thick quartz cover chip is positioned about 70~$\mu$m above the resonators. The cover chip shields the resonators from radiative heating and minimizes freezing of residual gas onto device surface during cooldown (see Appendix~\ref{Appendix:Surface_adsorption} for details), while minimally interfering with the microwave and optical probe. We investigated the effect of optical absorption heating from the probe laser. The incident laser introduced no more than 0.2~K temperature rise in the cryostat base temperature as measured by the silicon diode temperature sensor.  By tracking the dependence of Q and mode frequency on laser power and comparing these with their measured temperature dependence, we bound the temperature rise of the device due to laser heating to be less than 1~K.  In all, the base temperature of the setup, as measured by the diode thermometer, is limited to about 8~K by the opening in the radiation shield that is required for translating the microwave probe.  Since the temperature sensor is not located directly adjacent to the resonator, we cannot fully rule out the possibility that the actual local temperature of the resonator may be slightly higher due to thermal gradients.

\section{\label{Exp_results} Experimental results}
We characterize two resonator arrays, each comprising resonators with the defect width swept from 23.1~$\mu$m to 28.6~$\mu$m. One array has 5 Bragg mirror unit cells on both ends, while the other array has 3 Bragg mirror unit cells, referred to as 5M and 3M arrays, respectively. All resonators have the same mirror unit cell design. The temperature and pressure were varied to investigate their effects on the quality factor of the quartz PnC resonators.

\begin{figure*}
\centering
\includegraphics[width = \linewidth]{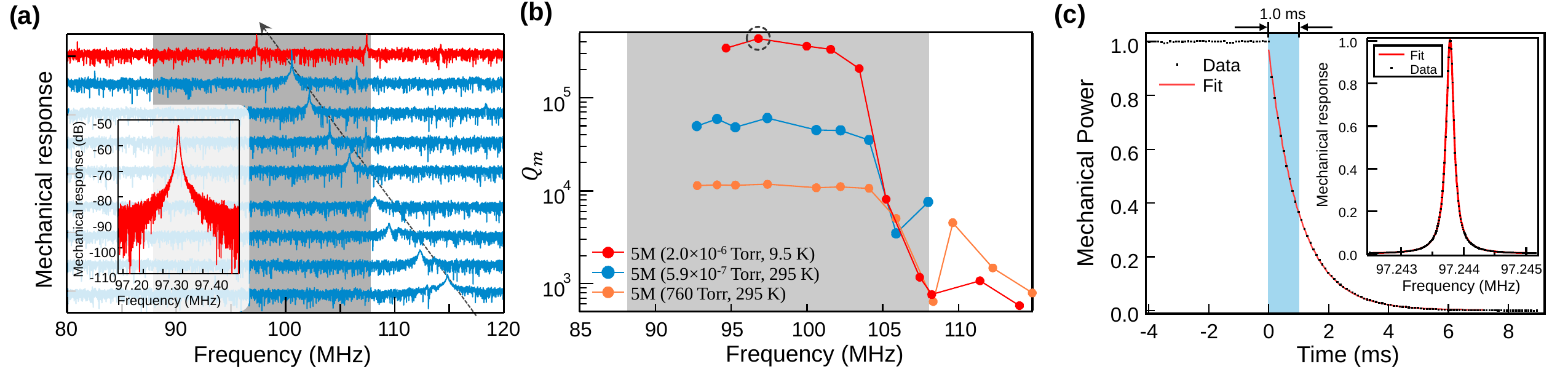}
\caption{\label{exp_results}\textbf{5M resonator array characterization.} (a) Mechanical response (arbitrary units) of the array of resonators with 5 PnC mirror unit cells and varying defect width at room temperature  and in vacuum.  Data traces are offset for clarity. The gray shaded area represents the simulated complete band gap of PnC mirror. The upper trace (red) corresponds to the resonator with the largest defect width, and the dotted line approximately tracks the frequency of the width-extension mode for devices of different width. The inset provides a zoom-in view of the red trace. (b) Measured quality factors of the resonator array at different temperatures and pressures. The max-Q mode is highlighted with a black circle. Statistical error bars are smaller than the size of the markers and not shown.  (c) Ringdown measurement of the max-Q mode at 13~K (red). An exponential fit (black) indicates a lifetime of 1.02~ms. The inset shows the corresponding mechanical frequency response.}
\end{figure*}

Figure~\ref{exp_results}(a) shows the optically detected mechanical response of the 5M resonator array at room temperature and a pressure of $2.0\times 10^{-6}$~Torr. We observe a single defect mode in our expected band gap region whose frequency decreases as the defect width increases. Our microwave-driven, optically detected measurement setup yields a high signal-to-noise ratio of up to 40~dB. The measured frequencies of the defect modes are within a few percent of the simulated values. By fitting the mechanical signal to a Lorentzian function, we extract the quality factor and frequency of the mechanical modes. During measurements, the microwave probe is positioned 10's~$\mu$m away from the resonator and is highly undercoupled, so the fitted Q closely approximates the intrinsic Q of the mechanical mode.

Figure~\ref{exp_results}(b) shows the measured Q's of the 5M array. There are several loss channels that determine the Q. One is anchor loss, which occurs when phonons leak into the environment through the resonator’s clamping points. The PnC Bragg mirror unit cells are designed to confine the modes and isolate them from the anchoring points, preventing acoustic energy from radiating into the bulk. To evaluate the efficacy of the PnC acoustic shielding, we track the Q as a function of the mode frequency. As shown in Fig.~\ref{exp_results}(b), as the mode frequency enters the acoustic band gap, Q increases. At room temperature and in vacuum, the Q of 5M array plateaus at around $5\times 10^4$ for mode frequencies near the center of the band gap. We observe a similar behavior in the 3M array with a similar maximum Q (see Appendix~\ref{Appendix:3M_array}). However, increasing the number of PnC mirror unit cells should exponentially reduce the anchor loss for modes in the band gap. These trends indicate that the PnC acoustic shielding highly suppresses the anchor losses for modes in the band gap, and internal loss mechanisms dominate the Q. These measurements are consistent with simulations, which show a maximum radiative Q $\sim 10^8$ (see Appendix~\ref{Appendix:3M_array}).

The plateau of the Q inside the band gap is a function of pressure and temperature. At room temperature and atmosphere pressure, the Q is limited to around $1\times10^4$ by gas damping \cite{schmid_fundamentals_2023}. At the high vacuum level we reach, gas damping is expected to be negligible. Cooling the resonator below 10~K raises the plateau about an order of magnitude to over $5\times10^5$ by suppressing the internal losses, as detailed in the next section. We focus further measurements on the mode with the maximum Q (max-Q mode). A ringdown measurement to characterize the lifetime of the max-Q mode at 13~K is shown in Fig.~\ref{exp_results}(c). The extracted mode lifetime is $1.023\,\pm\,0.005~\textrm{ms}$, consistent with the value derived from the frequency response, as shown in the inset.

For piezoelectric devices, electrodes also introduce losses when they are in direct contact with the resonators. Such losses include internal losses in polycrystalline metals \cite{TED_in_electrodes_2017}, ohmic losses \cite{Dissipation_analysis_2020, Loss_channel_Amir_2021}, surface and interfacial losses, as well as losses from two-level systems (discussed further below) \cite{Loss_channel_Amir_2021, emser_thin-film_2024}. Our current weakly coupled off-chip microwave probe does not induce significant losses. 
However, for our eventual quantum memory application, a stronger electro-mechanical coupling is required.  We present below, a design for optimized off-chip planar superconducting electrodes that provides comparable coupling rates to that achievable with directly contacted electrodes, while not introducing excess mechanical loss.

\section{\label{sec:level1} Loss mechanism}
Various external and internal dissipation mechanisms contribute to the energy dissipation of mechanical resonators, yielding a total quality factor ($Q_m$), as shown in Eq.~\ref{total_loss}.
\begin{align}\label{total_loss}
    \frac{1}{Q_m}&=\underbrace{\frac{1}{Q_{gas}}+\frac{1}{Q_{electrode}}+\frac{1}{Q_{anchor}}+\cdots}_{extrinsic}\notag\\
    &+\underbrace{\frac{1}{Q_{phonon-phonon}}+\frac{1}{Q_{TED,\,TLS,\,impurity}}+\dots}_{intrinsic}
\end{align}

In the previous section, we have shown that our device design effectively mitigates external loss mechanisms.  Intrinsic losses in crystalline solids include: 
(1) Phonon-phonon scattering, arising from anharmonic interactions between the acoustic waves and thermal phonons. At high temperature, when the acoustic wavelength is large compared to the mean free path of thermal phonons ($\omega_m\tau_{bath}\ll1$ where $\tau_{bath}$ is the thermal phonon relaxation time), such phonon-phonon scattering can be described as Akhiezer damping \cite{Anelasticity_of_solid_1972,Physical_acoustics_Vol_8}.  At low temperature ($\omega_m\tau_{bath}>1$), thermal phonons have a long mean free path and well-defined energy and momentum \cite{Physical_acoustics_Vol_8, cleland_foundations_2003}. In this regime, Landau-Rumer (L-R) theory models the acoustic mode as a beam of phonons interacting with ballistic thermal phonons. The resulting dissipation depends on the specific three-phonon scattering process and typically scales as $Q^{-1}_{L-R}\propto T^{n}$, with $n\approx 4$ and independent of $\omega_m$ \cite{Anelasticity_of_solid_1972,cleland_foundations_2003, rodriguez_direct_2019}. (2) Thermoelastic damping (TED), where strain induced by mechanical motion creates a temperature gradient via thermal expansion, causing irreversible diffusive heat flow~\cite{lifshitz_TED_2000, Quantum_Limit_of_Q_2013}. For the width-extension mode studied in this work, the strain distribution is relatively uniform with minimal gradient, hence the TED is expected to be small \cite{Chandorkar_Kenny_2008,Quantum_Limit_of_Q_2013}. There is some evidence that TED can dominate devices with directly contacted metal electrodes \cite{TED_in_electrodes_2017}, which are not present in our device. (3) Coupling to impurities such as alkaline and metal ions introduced during quartz growth or fabrication processes. Such impurities play an important role in the dissipation of macroscopic quartz resonators \cite{CRA_impurity_Zimmer_Tünnermann_2007}. The impurity can be modeled as two metastable states with transitions activated thermally \cite{Dissipation_in_NEMS_2014, schmid_fundamentals_2023}.  Akhiezer damping, TED, and impurity ion couplings can be modeled with Zener's approach as \cite{cleland_foundations_2003,schmid_fundamentals_2023}:
\begin{align}\label{TED_Akhiezer}
    Q^{-1}(\omega, T)=\Delta\frac{\omega\tau_{bath}}{1+(\omega\tau_{bath})^2}
\end{align}
featuring a Debye peak around $\omega\tau_{bath}=1$, with dimensionless parameter $\Delta$ representing the relaxation strength of a given loss mechanism. Generally $\tau_{bath}$ decreases exponentially with $T$, leading to a temperature dependence of Q.

\begin{figure}
    \centering
    \includegraphics[width=\linewidth]{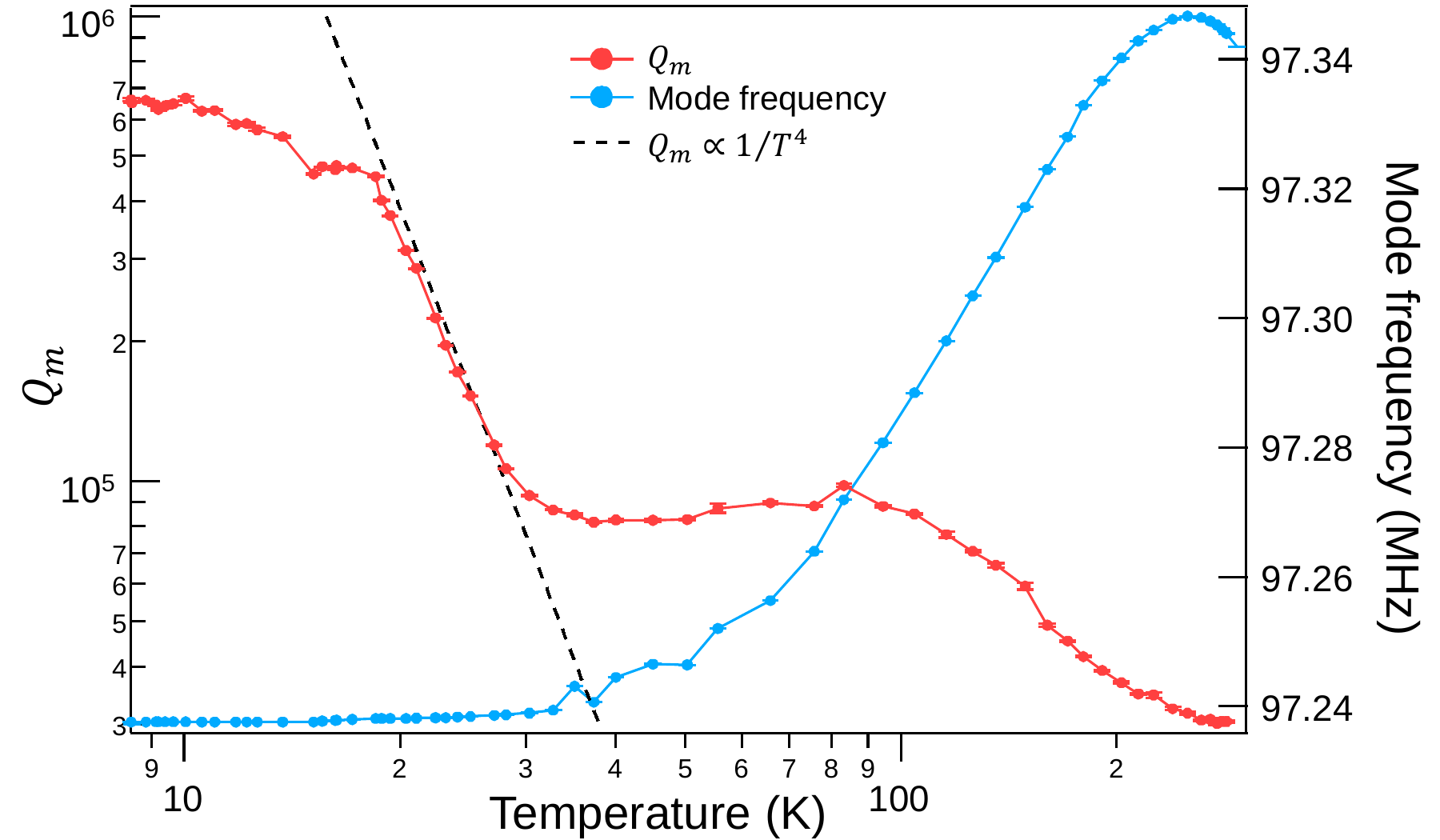}
    \caption{\textbf{Temperature dependence measurements.} Q factor and mode frequency of the max-Q mode measured at different temperatures. Also shown is a $T^4$ power law characteristic of Landau-Rumer limit of phonon-phonon scattering.}
    \label{T_dependence}
\end{figure}

Figure~\ref{T_dependence} shows the temperature dependence of both the quality factor and mode frequency of the max-Q mode.  As the max-Q mode is cooled from 300~K to around 8~K, its quality factor improves by more than one order of magnitude. The Q roughly plateaus from around 100~K to 30~K and shows a sharp increase below 30~K. This can likely be explained by the transition of phonon-phonon scattering loss from Akhiezer damping to Landau-Rumer damping. The loss peaks at around 40~K where $\omega\tau_{bath}\approx1$ (see Eq.~\ref{TED_Akhiezer}). For a range below 30~K, the $Q^{-1}$ exhibits a roughly $T^4$ Landau-Rumer-type dependence as shown with the dashed line \cite{Anelasticity_of_solid_1972}.
Below 20~K the Q increases more slowly with sublinear temperature dependence and reaches a maximum value of approximately $6.8\times10^5$ at around 8~K.  In this temperature range, other groups have attributed such a slow increase in Q to be dominated by TLS losses \cite{Quartz_TLS_loss_2007, Quartz_loss_Galliou_Tobar_Ivanov_2014}. Furthermore, as mentioned in the experimental setup section, the actual local resonator temperature may deviate from the sensor reading, potentially partially accounting for the reduced rate of Q enhancement at the lowest temperatures.

At millikelvin temperatures and low phonon occupation where we expect our resonator to perform as a quantum memory, most loss mechanisms freeze out, leaving two-level system loss the dominant loss mechanism~\cite{Loss_channel_Amir_2021, emser_thin-film_2024,Gruenke-Freudenstein2025}. TLSs are defects deviating the device materials from ideal crystals. Such defects are inherent to amorphous materials but can also present in crystalline materials as atomic and molecular impurities or dislocations, and surface and interface states~\cite{Intrisinc_dissipation_Mohanty_Roukes_2002, TLS_Behunin_Rakich_2016, TLS_Müller_Cole_Lisenfeld_2019}. These defects can be modeled as an ensemble of two-level tunneling systems, with each TLS representing two eigenstates arising from atoms or atomic groups residing in an asymmetric double-well potential with tunneling transitions \cite{Low_temp_physics_2005, Ultralong_phonon_lifetime_2020}. Energy relaxation and dephasing of acoustic phonons can occur via either resonant or non-resonant coupling with TLSs. In the resonant case, TLSs directly absorb phonons from the acoustic wave. In the non-resonant case, the acoustic strain field deforms the TLS potential and cause energy dissipation by driving the TLS out of equilibrium. The TLS then relaxes back to thermal equilibrium, releasing energy as heat. TLSs are saturated at high temperatures and contribute minimally to the decoherence because they will not cause net energy flow. However, at low temperatures (typically below $\sim$10~K for silica glass \cite{TLS_in_quartz_1986, Damping_in_class_II_2005}), as other loss mechanism are suppressed, TLSs are unsaturated and are significant sources of decoherence \cite{TLS_Jäckle_1972}. In implementations where electrodes directly contact the resonators, TLSs losses at the electrode-dielectric interface are found to be the main loss contribution below 1~K \cite{Loss_channel_Amir_2021, emser_thin-film_2024}. Our non-contact superconducting electrode implementation is expected to avoid such TLSs.

Additionally, during cryogenic measurements, we observe the adsorption of water and other trace contaminants on the resonator surface (see details in Appendix~\ref{Appendix:Surface_adsorption}) \cite{Loss_channel_Amir_2021}. Though the 50~$\mu$m cover chip largely mitigates this effect, residual adsorbents may still persist and contribute to mechanical dissipation, limiting the maximum Q. Mass loading from such adsorption could also explain the anomalous reduction in mode frequency observed as the temperature decreases, which contrasts with the expected frequency increase resulting from material stiffening at lower temperatures \cite{Tarumi2007}.

\section{Coupling to superconducting qubits}
\begin{figure}[H]
\includegraphics[width = \linewidth]{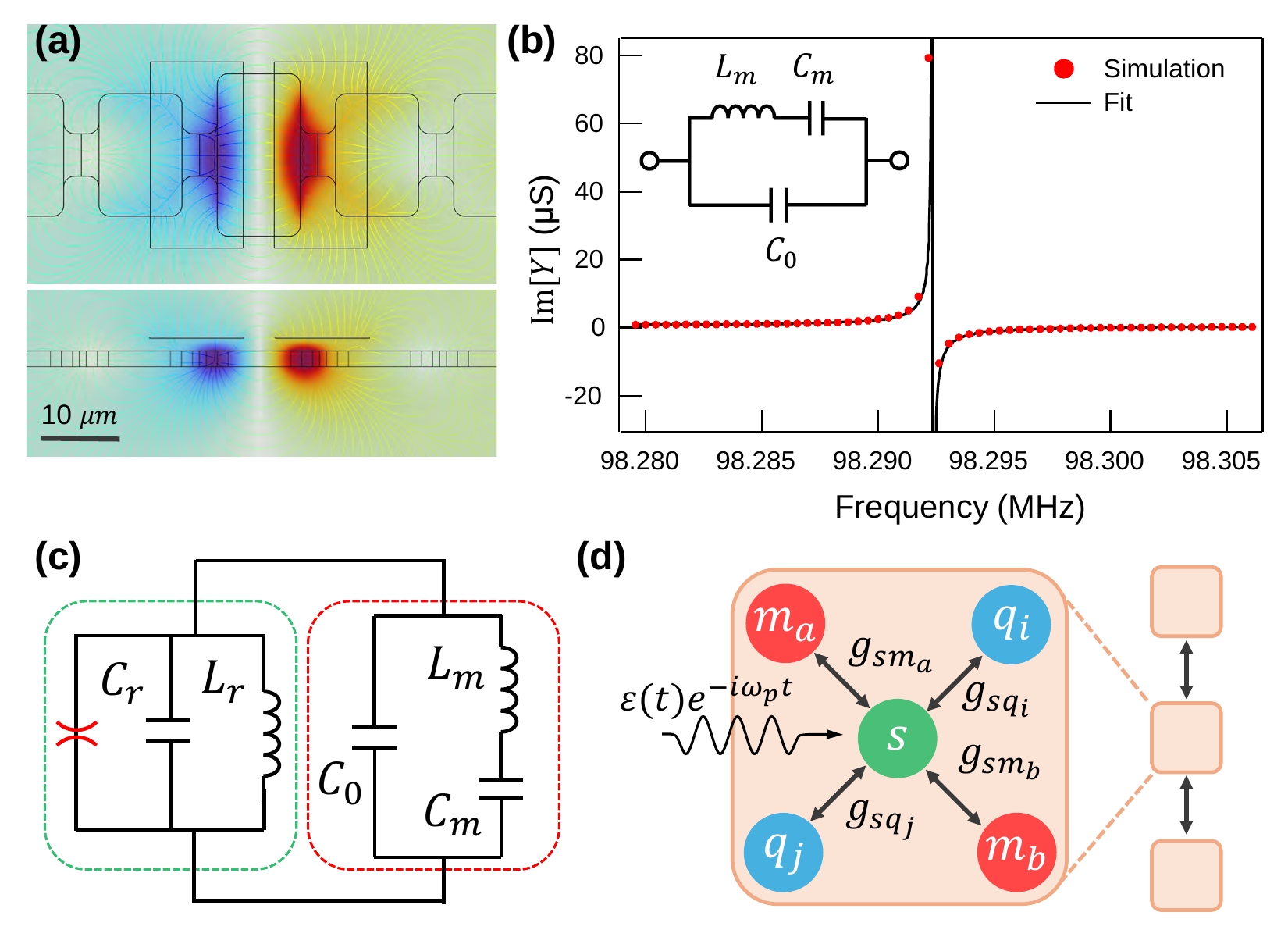}
\caption{\label{Coupling_scheme} \textbf{Hybrid acoustic quantum memory scheme.} (a) Top view and side view of electrical potential distribution of the width-extension mode, with the gray scale bar of 10~$\mu$m. Electrodes are positioned 1~$\mu$m above the defect region. (b) Simulated imaginary part of the admittance (Y) between the electrode pair as a function of frequency (red dotted curve), with the black curve showing a fit to the BVD model (inset). (c) Circuit diagram representing a mechanical mode (red dashed region) coupled to a nonlinear superconducting circuit (green dashed region) with its nonlinearity indicated by a red cross. (d) Vision of hybrid acoustic quantum memory, where mechanical mode (red circle) is parametrically coupled to qubits (blue circles) via SNAIL (green circle). This architecture is extendable to multiple modules (peach boxes) for building a scalable network.}
\end{figure}

The width-extension mode exhibits a well-defined electric potential distribution, indicating significant potential for strong piezoelectric coupling to other systems. To harness this coupling, we design non-contact electrodes on another chip as in Fig.~\ref{Coupling_scheme}(b), which will also contain the superconducting circuit system. To evaluate the mode's performance in coupling to other systems (e.g., superconducting qubit, SNAIL, SQUID~\cite{Goryachev2014, QC_catCode_Chamberland_2022}), we adopt the lossless Butterworth-Van-Dyke (BVD) equivalent circuit model as in Fig.~\ref{Coupling_scheme}. In this model, $C_0$ represents the static capacitance of the system, $C_m$ and $L_m$ correspond to the motional inductance and capacitance describing the mechanical mode. This mechanical mode is directly coupled to a nonlinear circuit modeled as an LC oscillator ($L_r$, $C_r$) in parallel with a nonlinear element. The coupling rate $g$ for their interaction when $C_r\gg C_0$, $C_m$ can be approximated as \cite{Loss_channel_Amir_2021}:
\begin{align}\label{gsm}
    g_{sm}\approx \frac{1}{2}\sqrt{\omega_r \omega_m}\,\sqrt{\frac{C_m}{C_r+C_m+C_0}}
\end{align}
with $\omega_r,\omega_m$ being the resonance frequency of the nonlinear circuit and the mechanical mode. Figure~\ref{Coupling_scheme}(b) shows the simulated imaginary part of the admittance between 
the a pair of rectangular electrodes straddling the central defect, positioned 1~$\mu$m above the resonator.  The simulation yields the equivalent circuit parameters for the dimensions of the max-Q mode: $C_0 = 8.96\times 10^{-16}$~F, $C_m = 1.38\times 10^{-19}$~F, $L_m = 18.9$~H (detailed in Appendix~\ref{Appendix:Electro_mechanical_coupling}). 

This resonator can couple to qubits, either resonantly or parametrically, with mechanical modes acting as quantum memory units. For example, we may resonantly couple the max-Q mode to a fluxonium qubit with a transition frequency of $\omega_r\approx2\pi\times$100~MHz and an equivalent capacitance $C_r\approx$ 30~fF \cite{Strong_dispersive_coupling_with_fluxonium_2023,Couple_to_heavy_fluxonium_qubit_2024}. From Eq.~\ref{gsm} we estimate a coupling rate $\frac{g}{2\pi}\approx$ 100~kHz leading to a $\frac{g}{\omega}\approx 0.1\%$. Although this value is lower than can be achieved with contacted electrodes on a lithium niobate mechanical resonator \cite{Strong_dispersive_coupling_with_fluxonium_2023},
it offers the advantage of contactless electrodes and low loss materials. We note that simulations of electrodes with zero air gap (e.g.~contacted electrodes) yields only a $\sim40\%$ increase in coupling rate. For most quantum memory applications, the benefits of contactless electrodes in reducing dissipation outweigh the modest reduction in coupling rate \cite{Loss_channel_Amir_2021, Noncontact_coupling_2024}.

Mechanical modes can also be parametrically coupled to superconducting qubits via three-wave mixing, through geometric capacitive coupling \cite{teufel2011,Alkim_Mohammad_2023, Couple_to_heavy_fluxonium_qubit_2024}, or mediated by a nonlinear circuit element like a SQUID 
\cite{QC_catCode_Chamberland_2022} or a SNAIL \cite{3-wave_mixing_SNAIL_Devoret_2017}.  Following the approach in \cite{Co_design_quantumModules, Chao_router_2023}, we propose a SNAIL-based architecture for coupling mechanical modes to transmon qubits with tunable coupling strength, enabling their operation as a quantum memory, as shown in Fig.~\ref{Coupling_scheme}(d). The SNAIL is a nonlinear circuit element flux-biased to exhibit strong third-order nonlinearity while suppressing even-order nonlinearities, thereby enabling efficient three-wave mixing \cite{3-wave_mixing_SNAIL_Devoret_2017,Chao_router_2023}. The ultimate system would consist of multiple interconnected modules, each containing a central SNAIL that couples to multiple mechanical modes and multiple transmon qubits. Such mechanical-SNAIL-qubit architecture offers a scalable, modular platform. The flexibility of control afforded by a strong parametric pump driving the SNAIL, allows the interactions be turned on and off with high fidelity,  thus preserving the coherence of the memory and mitigating unwanted hybridization during storage. 

To realize the basic building block of this architecture, a die supporting mechanical resonators will be flip-chip bonded~\cite{Flip-chip_2019, Malik2023} to another die housing the electrodes and SNAIL (see Appendix~\ref{Appendix:Electro_mechanical_coupling} for details). The combined SNAIL-mechanical assembly can be antenna-coupled to multiple transmons, with modules communicating via electromagnetic modes \cite{Co_design_quantumModules}. Choosing the frequencies of all elements in a module to be separated by several GHz allows low loss, low crosstalk coupling by driving the SNAIL at the difference frequency between any qubit and any mechanical mode $\omega_p=|\omega_{q}-\omega_m|$.  Here, the SNAIL is highly overcoupled to the pumping port to facilitate strong parametric driving, thus reducing its coherence time.  The SNAIL frequency is then chosen to be far off resonance with the pump to enable strong mechanics -- qubit coupling without populating the relatively lossy SNAIL mode and to allow for coupling to qubits with a wide range of frequencies. The resulting effective interaction Hamiltonian is:
\begin{align}\label{Interaction_Hamiltonian}
   \hat{H}^{eff}_{mq}=\hbar\left( g^{eff}_{mq}\hat{m}^\dagger\hat{q} + g^{eff*}_{mq}\hat{m}\hat{q}^\dagger\right)
\end{align}
with effective coupling strength:
\begin{align}
    \label{m-c coupling}
    g^{eff}_{mq}&\approx\eta\,6g_3\left(\frac{g}{\Delta}\right)_{ms}\left(\frac{g}{\Delta}\right)_{sq}
\end{align}
Here, $g_3$ is the strength of SNAIL's third-order nonlinearity and $\eta=\sqrt{n_s}e^{i\Phi_p}$ is proportional to the pumping strength, where $n_s$ is the occupancy of the SNAIL and $\Phi_p$ is the phase of pump signal. The factor $\left(\frac{g}{\Delta}\right)_{ij}$ denotes the ratio of the coupling rate to the detuning between components $i$ and $j$, with detuning $\Delta_{ij}\equiv \omega_i-\omega_j$ (details presented in Appendix~\ref{Appendix:Electro_mechanical_coupling}) \cite{Chao_router_2023}.

With an interaction Hamiltonian of the form in Eq.~\ref{Interaction_Hamiltonian}, tuning the pumping duration applied to the SNAIL can implement an iSWAP-like gate (detailed in Appendix~\ref{Appendix:Electro_mechanical_coupling}), allowing the coherent exchange of quantum information between the qubit and the mechanical mode \cite{Chao_router_2023}.  This operation can be used to write a state from the qubit into the mechanical memory, assuming that the mechanics is initially prepared in its ground state, and to subsequently read the stored state from the mechanics back into the qubit.  The corresponding swap time is given by $T_{\textrm{iSWAP}}=\frac{1}{2}\times\frac{2\pi}{g^{eff}_{mq}}$.
Given a SNAIL with resonance frequency about 1.13~GHz, characterized by $C_r = 0.26\textrm{~pF}, L_r = 75\textrm{~nH}$,  the coupling strength between the max-Q mode and the SNAIL is estimated as $g_{sm}\approx 2\pi\times 120~\textrm{kHz}$. With typical parameters $n_s\approx10, \left(\frac{g}{\Delta}\right)_{sq}\approx0.1, g_3=2\pi\times\textrm{100~MHz}$ \cite{Chao_router_2023}, the effective qubit-mechanical coupling strength is estimated to be $g^{eff}_{mq}\approx2\pi \times \textrm{20~kHz}$, yielding a swap time of $T_{\textrm{iSWAP}}\approx \textrm{20~$\mu$s}$.

Although our current estimated coupling rate is modest compared to other implementations \cite{Strong_dispersive_coupling_with_fluxonium_2023,Alkim_Mohammad_2023}, there are viable strategies for improvement. As shown in Eq.~\ref{gsm}, when $C_r\gg C_0, C_m$, with a given SNAIL, the coupling rate $g_{sm}$ can be enhanced by increasing the motional-capacitance mode-frequency product $\omega_mC_m$. An effective strategy is to modify the central defect of the mechanical resonator into an array of multiple defect unit cells and proportionally enlarge the coupling electrodes. This approach increases the piezoelectric coupling area and scales the motional capacitance approximately as $C_m\propto N$ ($N$ is the number of defect unit cells) while leaving $\omega_p$ largely unchanged, thus boosting $g_{sm}$ and $g^{eff}_{mq}$ by a factor of $\sqrt{N}$. This approach would also improve the coupling to other circuit systems, including fluxonium, by a similar enhancement factor. In Appendix~\ref{Appendix:Improving_Coupling}, we analyze the coupling performance of a resonator with 10 defect unit cells, which yields three times larger effective coupling rate, at a cost of introducing additional spurious mechanical modes with weak piezoelectric coupling. In addition, reducing the SNAIL capacitance and increasing the pumping strength on the SNAIL can further enhance the effective coupling.  With these combined improvements, we expect to reach an effective coupling rate of over 100~kHz.

\section{\label{sec:level1} Conclusion and outlooks}
We have developed a fabrication process for making suspended thin-film quartz-on-silicon mechanical resonators at the 100~MHz, tens-of-micron scale. We also developed a piezoelectric-photoelastic measurement system to characterize arrays of 1-D quartz phononic crystal resonators. We demonstrate that the acoustic shielding of our phononic crystal allows us to reach the intrinsic mechanical loss limit for a width-extension mode. The measured quality factor was up to $6.8\times 10^5$ at 8~K, corresponding to a lifetime of 1.0~ms. We developed a design for contactless electrodes to piezoelectrically couple our devices to superconducting qubits. This design eliminates two-level system losses at metal-dielectric interfaces, which are common limiting sources of decoherence for low-temperature mechanical resonators. We propose a SNAIL-mediated scheme for parametrically coupling the mechanical modes to transmon qubits to serve as quantum memory units.

\begin{acknowledgments}
This work was partially funded by the AFOSR under Grant No. FA9550-21-1-0118. Work performed in the University of Pittsburgh Nanofabrication and Characterization Core Facility (RRID:SCR\_05124) and services and instruments used in this project were graciously supported, in part, by the University of Pittsburgh. We thank the NFCF staff for their assistance. Support for JR and MH was provided by the U.S. Department of Energy, Office of Science, National Quantum Information Science Research Centers, the Co-design Center for Quantum Advantage (C2QA) under Contract No. DE-SC0012704.
\end{acknowledgments}

\appendix
\section{Device fabrication}\label{Appendix:Device_fabrication}
Due to difficulty sourcing commercial thin-quartz-on-silicon wafers, we start the fabrication process with a 2-inch diameter 70~um thick Z-cut quartz wafer (Precision Micro-Optics) bonded onto a 4-inch high-resistivity silicon wafer (El-Cat. Inc.) via commercial atomic diffusion bonding (Partow Technologies). The bonded wafer was then diced into small dies using an automatic dicing saw (DAD 321)  for subsequent processing.

We thinned the bonded dies using ICP-RIE (Plasma-Therm Apex ICP RIE) with the recipe \{SF6: 15~sccm, Ar: 45~sccm, pressure: 9.8~mTorr, Bias: 200~W, Temp set point: $25^\circ $C\}. The 1:3 SF6-to-Ar ratio was chosen to suppress the micromasking effect \cite{Optimal_etch_SF6_Ar_2013, Deep_RIE_quartz_2020}. Sapphire handle wafers were used to carry the bonded dies because sapphire has a high resistance to fluoride-based etching. The ICP Power was initially set at 800~W, which etches quartz at a rate of about 0.3~$\mu$m/min. To minimize loading effects \cite{Loading_effect_2014} and improve etch uniformity, we surrounded the bonded die with bulk quartz dies of similar thickness. To mitigate heating during the etching process, we paused every 10 minutes of etching to allow the system to cool down, with both bias and ICP power turned off. As we approach the target quartz thickness, we reduced the ICP power to 100~W, yielding an etch rate of approximately 0.065~$\mu$m/min for better thickness control. The thickness of the quartz layer was monitored with a reflectometer (Filmetrics F40 Thin-Film Analyzer), an optical profilometer (Bruker Contour GT-I) and a SEM (FEI Scios Dual Beam System, ThermoFisher Apreo HiVac SEM).

When the quartz layer was thinned to 4.5~$\mu$m, we characterized its surface roughness with AFM, exhibiting a mean roughness of 0.7~nm. We then patterned circular nickel masks at the corners of each die using photolithography (MLA100), e-beam evaporation (Plassys MEB550S) and lift-off.  We etched the quartz layer with ICP-RIE for another 1~$\mu$m, which formed 1~$\mu$m-tall quartz pillars used to define the air gap thickness for a later flip-chip bonding process. The height of the pillars was measured with a surface profiler (Bruker Dektak-XT). After these etching steps, the quartz layer has been thinned to the target thickness of 3.5~$\mu$m. We transferred the resonator pattern onto the quartz layer via photolithography and deposited a Ni mask. A subsequent ICP-RIE step etched through the quartz and into the silicon substrate in the area outside the resonator pattern. The Ni mask was then removed with Cr etchant (Transene 1020). Finally, the quartz resonators were released by undercutting the silicon beneath them with a 30\% KOH solution at 80$^\circ$C, followed by cleaning with IPA and DI water.  After the KOH etch, the silicon has been etched to a depth of 40~$\mu$m.

\begin{figure}[h]
    \centering
    \includegraphics[width=\linewidth]{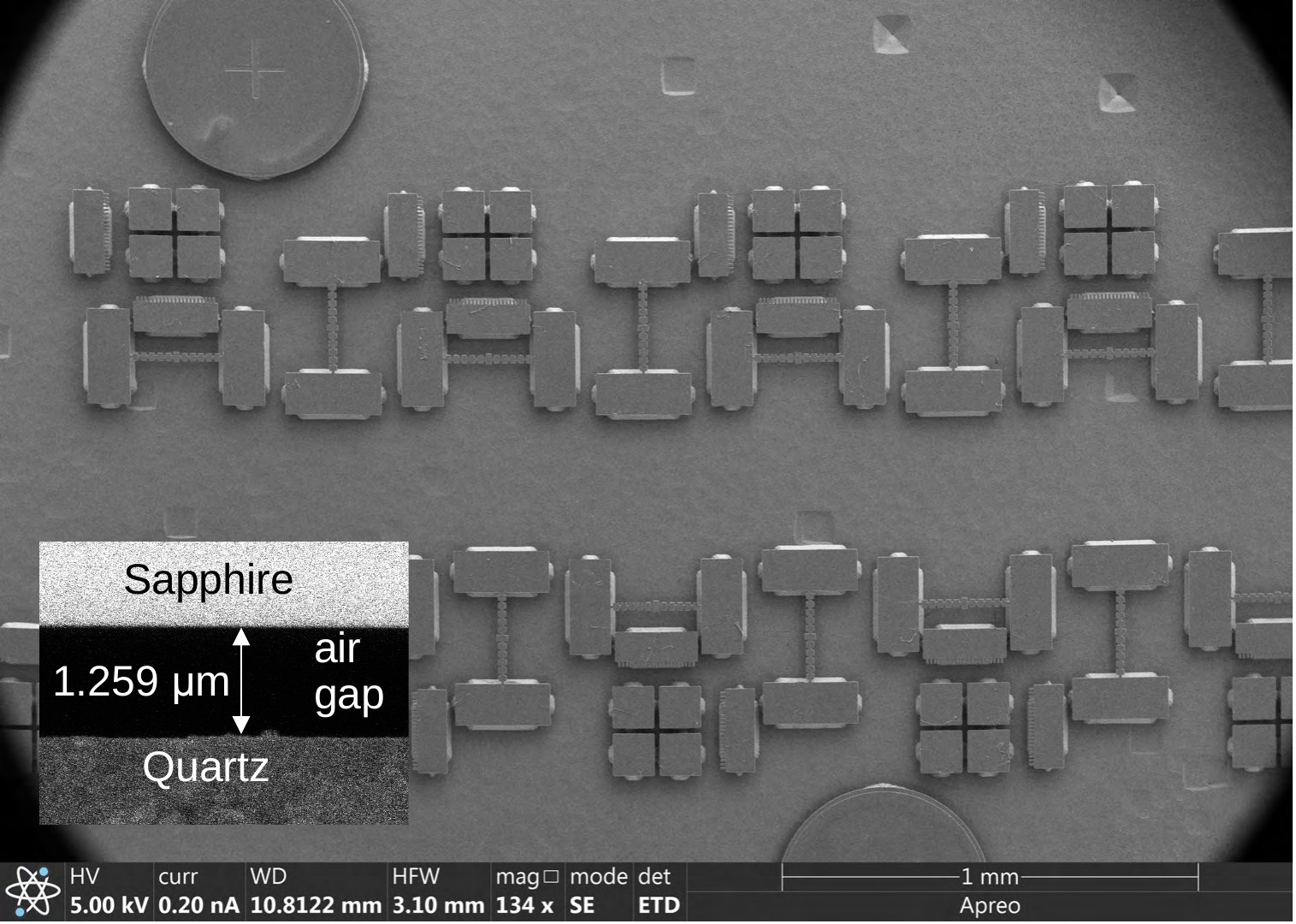}
    \caption{SEM image of a fabricated die with suspended quartz resonators. Quartz pillars for controlling the air gap during flip-chip bonding, and reticles including alignment crosses and vernier scales for chip aligning are also visible. Inset is an edge view of a mechanical resonator die flip-chip bonded to a sapphire die.}
    \label{63A_device_chip}
\end{figure}

Figure~\ref{63A_device_chip} shows an SEM image of a complete die. In addition to the resonators and their anchor pads, the circular pillars and alignment reticles for flip-chip bonding are visible. We performed a bonding test with a test sapphire chip using a home-built chip aligner and epoxy bonding. The inset shows a 1.3~$\mu$m air gap between a resonator chip with 1~$\mu$m pillars and a test sapphire chip. By imaging through the transparent sapphire chip, we achieved micron-level alignment tolerance for in-plane alignment utilizing the reticles.

\section{Electro-mechanical coupling}\label{Appendix:Electro_mechanical_coupling}

\begin{figure}[h]
    \centering
    \includegraphics[width=0.8\linewidth]{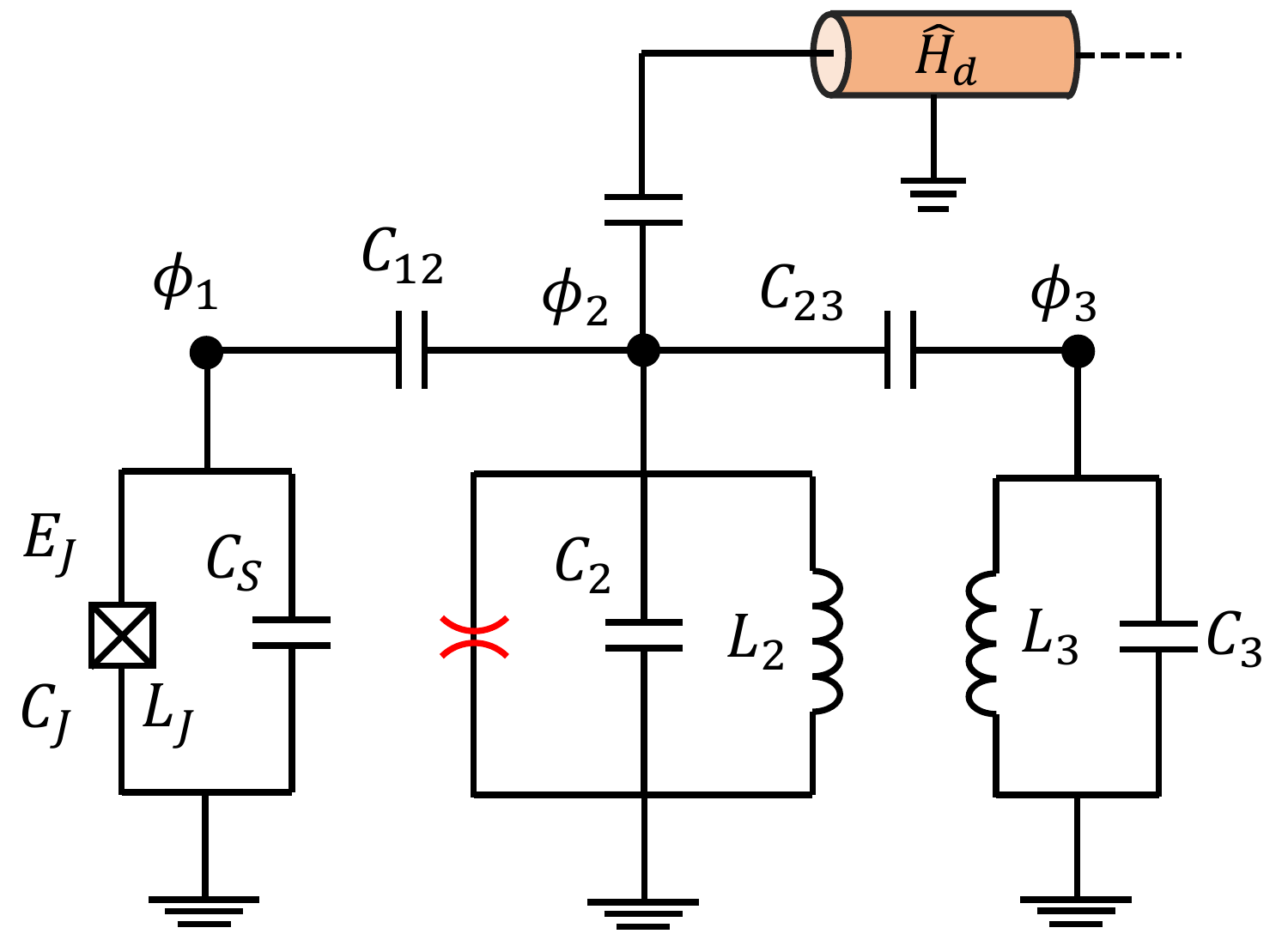}
    \caption{\textbf{Circuit model of Transmon-SNAIL-Mechanical mode system.} Each component is represented with a corresponding equivalent circuit. $\phi_i$ denotes the flux at each node, and $C_{ij}$ represents the coupling capacitance between components. The cross box represents the Josephson junction in the tramsmon, and the red cross indicates the nonlinear inductive element of the SNAIL. External flux is applied to capacitively drive the SNAIL, with $\hat{H}_d$ representing the driving term in the system Hamiltonian.}
    \label{Transmon_SNAIL_Mechanical}
\end{figure}

The complete Transmon-SNAIL-Mechanical resonator system is modeled as shown in Fig.~\ref{Transmon_SNAIL_Mechanical} where the SNAIL is driven by external flux. The transmon qubit is represented as a Josephson junction with inductance $L_J$ and capacitance $C_J$, shunted with a capacitor $C_S$, with total capacitance $C_1\equiv C_J+C_S$. The SNAIL is modeled as a harmonic LC oscillator ($L_2$, $C_2$) shunted with a nonlinear inductor, indicated in red. For convenience of analysis, we model the mechanical mode with an equivalent LC circuit model ($L_3$, $C_3$) \cite{Loss_channel_Amir_2021} rather than BVD model as in the main text. We use $C_{ij}$ to represent the coupling capacitance between mode $i$ and $j$.

The Hamiltonian of the system can be determined by circuit quantization following the derivations in \cite{Circuit_QED_Blais_Wallraff_2021,Chao_router_2023}. We begin by considering the static Hamiltonian in the absence of a driving signal applied to the SNAIL:
\begin{align}
    H_0&=H_L+H_{NL}\notag\\
    &=\hbar\omega_qq^\dagger q+\hbar \omega_ss^\dagger s+\hbar \omega_mm^\dagger m +\hbar g_{qs}(qs^\dagger+q^\dagger s)\notag\\
    &+\hbar g_{sm}(sm^\dagger+s^\dagger m)\underbrace{-\frac{E_C}{12}(q+q^\dagger)^4}_{H^Q_{NL}}+\underbrace{\hbar g_3(s+s^\dagger)^3}_{H^S_{NL}}\\
    g_{ij}&=\frac{1}{2}\sqrt{\omega_i\omega_j}\frac{C_{ij}}{\sqrt{(C_i+C_{ij})(C_j+C_{ij})}}
\end{align}
Here, $q$, $s$, $m$ represent annihilation operators of the qubit, SNAIL, mechanical mode, respectively. The first five terms represent the linear parts of the Hamiltonian with $g_{ij}$ representing the linear coupling rate between $i,j$. The fourth term $H^Q_{NL}$ accounts for transmon's nonlinearity with anharmonicity $-E_C\equiv-\frac{e^2}{2(C_J+C_S)}$, allowing the transmon to be treated as a two-level system where we can make the replacements $q\rightarrow\sigma_-=\ket{g}\bra{e}, q^\dagger\rightarrow\sigma_+=\ket{e}\bra{g}$. The last term $H^S_{NL}$ captures the 3rd-order nonlinearity of SNAIL. 

Treating the nonlinear terms as perturbations, we rewrite the Hamiltonian in the eigenbasis of the linear (harmonic) part $H_L$. In the weak dispersive regime where $|\lambda_{ij}|\ll1$ with $\lambda_{ij}\equiv\frac{g_{ij}}{\omega_i-\omega_j}$, we diagonalize the linear part using perturbation methods to get the new basis (dressed eigenstates) with corresponding dressed operators and dressed frequencies:
\begin{alignat}{2}\label{dressed states}
    \begin{cases}
      \hat{q}^\prime=\hat{q}+\lambda_{qs}\hat{s}\\
    \hat{s}^\prime=\hat{s}-\lambda_{qs}\hat{q}-\lambda_{sm}\hat{m}\\
    \hat{m}^\prime=\hat{m}+\lambda_{sm}\hat{s}  
    \end{cases}
    \begin{cases}
        \omega^\prime_q=\omega_q+\lambda_{qs}g_{qs}\\
    \omega^\prime_s=\omega_s-\lambda_{qs}g_{qs}-\lambda_{sm}g_{sm}\\
    \omega^\prime_m=\omega_m+\lambda_{sm}g_{sm}
    \end{cases}
\end{alignat}
Now $H_L\approx\hbar\omega^\prime_qq^{\prime^\dagger}q^\prime+\hbar\omega^\prime_ss^{\prime^\dagger}s^\prime+\hbar\omega^\prime_mm^{\prime^\dagger}m^\prime$. $H^S_{NL}$ can be written as: $H^S_{NL}=\hbar g_3[(s^\prime+\lambda_{qs}q^\prime+\lambda_{ms}m^\prime)+(s^{\prime^\dagger}+\lambda^*_{qs}q^{\prime^\dagger}+\lambda^*_{ms}m^{\prime^\dagger})]^3$ which includes the three-wave mixing terms $s^\prime m^\prime q^{\prime^\dagger}, s^{\prime^\dagger} m^{\prime^\dagger} q^\prime$.

As indicated in Eq.~\ref{dressed states}, modes $m$ and $s$ are weakly hybridized, potentially leading to the dressed mechanical mode $m^\prime$ inheriting excess dissipation from the low-Q SNAIL. The modified energy decay rate is given by: $\Gamma^\prime_m=\Gamma_m+\lambda^2_{sm}\Gamma_s$ where $\Gamma_i$ denotes the loss rate of system $i$. Given the numbers from the main text, $\lambda_{sm}\sim10^{-4}$, and $\Gamma_s\sim10^7~\textrm{s}^{-1}$, this hybridization limits the lifetime of the $m^\prime$ mode to $\sim$10~s. Hence, the lossy SNAIL won't cap the storage time of our current quantum memory.

Now we introduce the driving term on SNAIL to realize parametric coupling between qubit and mechanical mode:
\begin{align}
    H_d=\hbar\left(\epsilon(t)e^{-i\omega_dt}-\epsilon^*(t)e^{i\omega_dt}\right)(s^{\prime^\dagger}-s^{\prime})
\end{align}
where $\omega_d$ is the drive frequency and $\epsilon(t)$ sets the drive amplitude.
The total Hamiltonian is then given by $H=H_0+H_d$. To incorporate the effect of the driving signal on the evolution of the system, we apply a displacement transformation $\hat{D}(\alpha)=e^{\alpha s^\dagger-\alpha^*s}$ with $\alpha=\frac{\epsilon(t)}{\omega_s-\omega_d}e^{-i\omega_dt}-\frac{\epsilon^*(t)}{\omega_s+\omega_d}e^{i\omega_dt}$. The transformed Hamiltonian is:
\begin{align}
    H^D&=D(\alpha)(H_0+H_d)D^\dagger(\alpha)\notag\\    &=\underbrace{\hbar\omega^\prime_qq^{\prime^\dagger}q^\prime+\hbar\omega^\prime_ss^{\prime^\dagger}s^\prime+\hbar\omega^\prime_mm^{\prime^\dagger}m^\prime}_{H^D_0}\notag\\
    &-\frac{E_C}{12}[(q^\prime-\lambda_{qs}s^\prime+\lambda_{qs}\eta e^{-i\omega_dt})+h.c.]^4\notag\\
    &+\hbar g_3[(s^\prime+\lambda_{qs}q^\prime+\lambda_{ms}m^\prime-\eta e^{-i\omega_d t})+h.c.]^3
\end{align}
with definition $\eta\equiv\frac{2\omega_d}{\omega^2_s-\omega^2_d}\epsilon(t)$ such that $|\eta|^2$ represents the effective photon occupation of the SNAIL mode.
Then we transform to the interaction picture under the rotating frame transformation $U_R=e^{\frac{iH^D_0t}{\hbar}}$. When driving the SNAIL at the difference frequency $\omega_d=\omega_q-\omega_m$, we can apply the rotating wave approximation (RWA), where we drop fast-oscillating terms that average to zero. We obtain an effective Hamiltonian containing only slowly rotating components (for simplicity, the prime symbols are dropped in the following derivations, assuming $|\lambda_{ij}|\ll1$) as:
\begin{align}\label{Hamiltonian_RWA}
    H^{eff}\approx &-\frac{E_C}{2}(q^\dagger qq^\dagger q+\lambda^4_{qs}s^\dagger ss^\dagger s)-2E_C\lambda^2_{qs}q^\dagger qs^\dagger s\notag\\
    &+\hbar g^{eff}_{qm}(qm^\dagger e^{-i\phi_d}+q^\dagger me^{i\phi_d})
\end{align}
The first term in the first bracket represents the self-Kerr effect of the qubit which yields its anharmonicity. Here we assume the qubit's anharmonicity is larger than mechanical mode frequency ($E_C/\hbar>\omega_m$) and so that we will not have frequency collisions (at least to the leading order) which excite the qubit beyond the $\ket{g}$, $\ket{e}$ manifold. The second term in the first bracket represents the SNAIL's anharmonicity, which is negligible given a small $\lambda_{qs}$. The third term is a cross-Kerr interaction term that accounts for the frequency shifts of the qubit and the SNAIL arising from their coupling. These first three terms originate from the qubit's Kerr nonlinearity \cite{Circuit_QED_Blais_Wallraff_2021}. The last two terms describe a beam-splitter-type interaction between the qubit and the mechanical mode, with an effective coupling rate $g^{eff}_{qm}\equiv -6g_3\lambda_{qs}\lambda_{ms}|\eta|$ and phase of driving signal $\phi_d\equiv \textrm{arg}(\eta)$. In our analysis, we assume the cross-Kerr term is small and focus on the beam-splitter interactions.

With a constant drive on the SNAIL, the time-evolution operator of the system is $U(t)=\textrm{exp}(-\frac{iH^{eff}t}{\hbar})$. The transmon's inherent anharmonicity permits its treatment as a two-level system. Prior to quantum information storage, the mechanical resonator will be initialized to its ground state via SNAIL-mediated sideband cooling \cite{Mechanical_cooling_Martin_et_al_Zoller_2004}.  Given this initialization, the coherent dynamics allows the mechanical mode to remain within the zero- and one- phonon Fock state manifold.  Setting the driving time as $t=\frac{\pi}{2g^{eff}_{qm}}$ and phase as $\phi_d=\pi$ realizes an iSWAP-like gate which writes the qubit state into the mechanical mode with swap time $T_{\textrm{iSWAP}}=\frac{1}{2}\times\frac{2\pi}{g^{eff}_{mq}}$.  Next, letting the system evolve with the pump off ($\eta=0$) decouples the mechanics, SNAIL and qubit, with the mechanical resonator storing the quantum state unperturbed.  The state stored in the mechanics can then be returned to the qubit with a subsequent iSWAP with an appropriately chosen $\phi_d$, and assuming that the qubit is in its ground state at the start of this read swap.

\section{Improving qubit-mechanical coupling rates via multi-defect mechanical modes}\label{Appendix:Improving_Coupling}
Here, we discuss methods to improve the coupling rate between the mechanical resonator and other systems. When $C_r\gg C_0,C_m$, Eq.~\ref{gsm} can be simplified as $g\approx \frac{1}{2}\sqrt{\omega_r \omega_p}\,\sqrt{\frac{C_m}{C_r}}$. For a given $\omega_r,C_r$, improving the coupling rate requires improving the $\omega_pC_m$ product. The motional capacitance $C_m$ scales with the mode area for our planar thin-film devices. Hence, increasing the mode area can enhance the coupling rate. A straightforward idea is to uniformly scale up all of the resonator dimensions. However, this also reduces the frequency of the resonator, leaving the $\omega_pC_m$ product approximately constant. Since the mode frequency is primarily determined by the width of the defect, one might consider increasing the defect length (i.e. the dimension along the X-axis) while keeping its other dimensions fixed. However, increasing this dimension introduces a complicated transverse-mode structure, which reduces the piezoelectric coupling. One effective approach is to construct a mechanical mode by coupling an array of defect unit cells and employ the mode where all units oscillate in-phase as a collective width-extension mode. As shown in Fig.~\ref{Coupling_scheme}(a), due to the asymmetric electric potential distribution, such an in-phase mode would exhibit alternating regions of positive and negative potential. Efficient coupling to this mode requires design of interdigital electrodes similar to those used in SAW devices \cite{SAW_IDT_White_Voltmer_1965}, which will lower the tolerance for later chip-to-chip alignment and air gap.

\begin{figure}
    \centering
    \includegraphics[width=\linewidth]{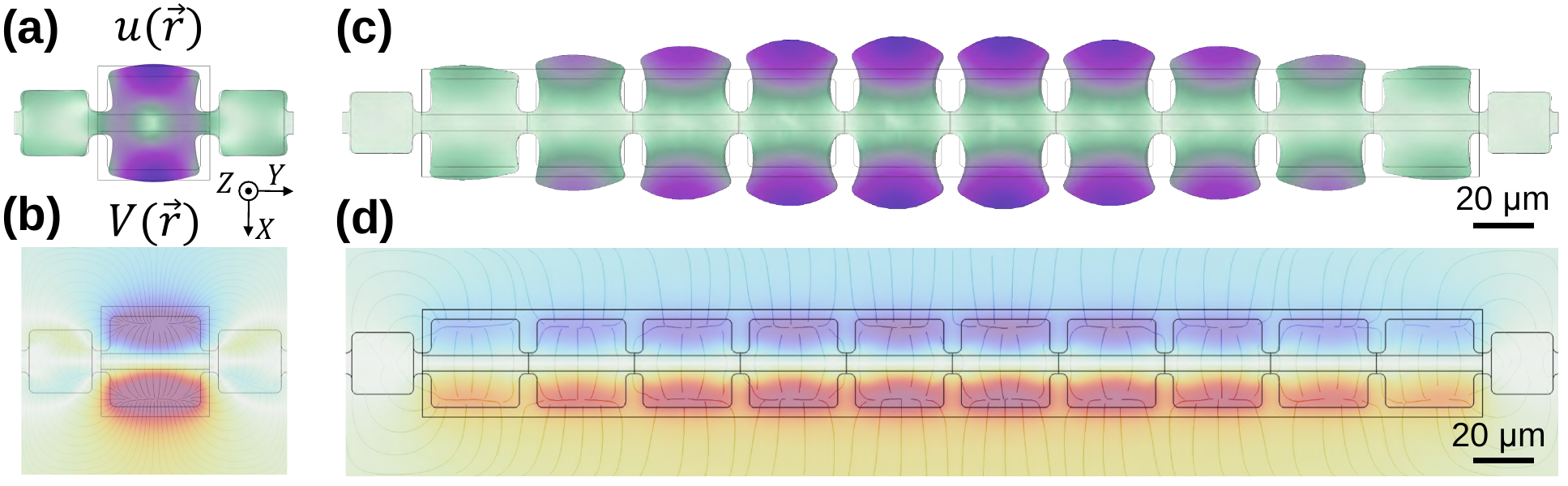}
    \caption{\textbf{Resonator with multiple defect units.} (a(b)): Mechanical displacement (Electrical potential) distribution of a Y chain resonator with a single defect unit. (c(d)): Mechanical displacement (Electrical potential) distribution of Y chain resonator with ten defect units. The spatial axes represent quartz crystallographic axes.}
    \label{Multi_defect_Y_chain}
\end{figure}

In addition to the resonators discussed in the main text where the phononic crystal chain structure is along the crystal X-axis (later referred to as X chains), we also fabricated and measured resonators of the same dimensions but along the quartz crystal Y-axis (Y chains). The Y chains behave similarly to the X chains, with the primary distinction being the orientation of the electric potential distribution, as shown in Fig.~\ref{Multi_defect_Y_chain}(a,b). The in-phase collective mode of the Y chain with multiple defect units exhibits two relatively uniform potential regions. This configuration enables a straightforward extension of the single-defect resonator electrode design to the multi-defect collective mode case, eliminating the need for the tight tolerance IDTs required for X chains. This design scales the piezoelectric coupling area and the motional capacitance approximately as $C_m\propto N$ ($N$ is the number of defect unit cells) while leaving $\omega_p$ largely unchanged, thus boosting $g_{sm}$ and $g^{eff}_{mq}$ by a factor of $\sqrt{N}$. Optimizing the geometry, we find that square unit cells (width equal to length) provide relatively wide band gaps (about 26\%) and optimize piezoelectric coupling rates.  Figure~\ref{Multi_defect_Y_chain} shows the simulated resonator with optimized dimensions and 10 defect units, possessing an in-phase collective width-extension mode at $96.8\textrm{~MHz}$. The simulated coupling rate of this mode to fluxonium with an equivalent capacitance $C_r=30~\textrm{fF},\omega_r=96.8\textrm{~MHz}$ is about 290~kHz, improved from the 104~kHz for single-defect mode. The parametric coupling rate to the transmon is enhanced to $g^{eff}_{mq}=2\pi \times \textrm{70~kHz}$, about triple the value for the single-defect mode.

\section{Photoelastic characterization}\label{Appendix:Photoelastic_characterization}
With an applied strain field, the change in the refractive index of quartz can be described by \cite{Photoelastic_effect}:
\begin{align}\label{photoelastic_effect}
    \Delta\left(\frac{1}{n^2}\right)_{ij}=p_{ijkl}S_{kl}
\end{align}
and the index ellipsoid can be written as (in the frame of quartz crystal orientation):
\begin{align} \label{Index ellipsoid}
    &x^2(\frac{1}{n_x^2}+\sum_{i=1}^6p_{1i}S_i)+y^2(\frac{1}{n_y^2}+\sum_{i=1}^6p_{2i}S_i)\notag\\
    &+z^2(\frac{1}{n_z^2}+\sum_{i=1}^6p_{3i}S_i)+2yz\sum_{i=1}^6p_{4i}S_i\notag\\
    &+2zx\sum_{i=1}^6p_{5i}S_i+2xy\sum_{i=1}^6p_{6i}S_i
    =1
\end{align}
where $S_{kl}$ are the components of the strain tensor, $p_{ijkl}$ are the photoelastic coefficients of quartz (in Eq.~\ref{Index ellipsoid} $S_i$, $p_{ij}$ are the strain component and the photoelastic coefficient, respectively, in Voigt notation):
\begin{align}
    p = 
    \begin{bmatrix}
    p_{11} & p_{12} & p_{13} & p_{14} & 0 & 0\\
    p_{12} & p_{11} & p_{13} & -p_{14} & 0 & 0\\
    p_{31} & p_{31} & p_{33} & 0 & 0 & 0\\
    p_{41} & -p_{41} & 0 & p_{44} & 0 & 0\\
    0 & 0 & 0 & 0 & p_{44} & p_{41}\\
    0 & 0 & 0 & 0 & p_{14} & \frac{p_{11}-p_{12}}{2}\\
    \end{bmatrix}
\end{align}
with $p_{11}$\,=\,0.16; $p_{12}$\,=\,0.27; $p_{13}$\,=\,0.27; $p_{14}$\,=\,-0.03; $p_{31}$\,=\,0.29; $p_{41}$\,=\,0.10; $p_{33}$\,=\,-0.047; $p_{44}$\,=\,-0.079 \cite{Narasimhamurty_1969, Detraux_Gonze_2001}.

For our width-extension mode, the dominant strain is $S_2 (S_{yy})$. Equation~\ref{Index ellipsoid} can be simplified as:
\begin{align}\label{index_ellipsoid}
    &x^2(\frac{1}{n_o^2}+p_{12}S_{yy})+
    y^2(\frac{1}{n_o^2}+p_{11}S_{yy})+
    z^2(\frac{1}{n_e^2}+p_{31}S_{yy})\notag\\
    &+2yz\left(-p_{41}S_{yy}\right)=1
\end{align}
with $n_o$=1.528 and $n_e$=1.536, the ordinary and extraordinary indices of unstrained quartz.

We aim to transform to the principal-axis system, and this can be achieved by diagonalizing the last three terms of Eq.~\ref{index_ellipsoid} using a coordinate rotation in the Y-Z plane with the rotation angle $\theta$ satisfying:
\begin{align}
    \tan(2\theta)=\frac{-2p_{41}S_{yy}}{\left(\frac{1}{n_o^2}+p_{11}S_{yy}\right)-\left(\frac{1}{n_e^2}+p_{31}S_{yy}\right)}
\end{align}
Given that $S_{yy}$ is small and quartz is birefringent, the required rotation angle is small. Thus, the principal index ellipsoid can be approximated as (here $x,y,z$ denote the principal axes) \cite{Photoelastic_effect, Atalar_Safavi-Naeini_Arbabian_2022}:
\begin{align}
    &x^2(\frac{1}{n_o^2}+p_{12}S_{yy})+
    y^2(\frac{1}{n_o^2}+p_{11}S_{yy})\notag\\
    &+z^2(\frac{1}{n_e^2}+p_{31}S_{yy})\approx1
\end{align}
The acoustic wave of the width-extension mode can be modeled as a standing wave with mechanical displacement distribution (here the mode is breathing along the crystal Y-axis, and the input laser is propagating along the crystal Z-axis through the resonator thickness):
\begin{align}
    u(y,t)=u_0\,\sin\left(\frac{\pi y}{L}\right)\cdot\sin\left(\omega_m t\right)
\end{align}
where $L$ is the defect width, equal to half of a mechanical wavelength, and $u_0$ is the amplitude of the mode.  The dominant strain is:
\begin{align}
    S_{yy}&=\frac{\partial u(y)}{\partial y}=u_0\,\frac{\pi}{L}\cos(\frac{\pi y}{L})\cdot\sin(\omega_m t)\notag\\
    &\equiv S_0\,\cos(\frac{\pi y}{L})\cdot\sin(\omega_m t)
\end{align}
and the principal indices of refraction can be approximated from Eq.~\ref{photoelastic_effect} as:
\begin{align}\label{refractive_index}
    \begin{cases}
    n_x=n_o-\frac{1}{2}n_o^3\,p_{12}\cdot S_0\,\cos(\frac{\pi y}{L})\cdot\sin(\omega_m t)\\
    n_y=n_o-\frac{1}{2}n_o^3\,p_{11}\cdot S_0\,\cos(\frac{\pi y}{L})\cdot\sin(\omega_m t)\\
    n_z=n_e-\frac{1}{2}n_e^3\,p_{31}\cdot S_0\,\cos(\frac{\pi y}{L})\cdot\sin(\omega_m t)
    \end{cases}
\end{align}
where the second term of each equation represents the photoelastic modulation. 

 We treat the optical signal collected by photodetector as interference between the laser reflected from the front and back surfaces of the resonator.  Given that the silicon surface under the quartz resonator is rough and angled after the KOH etch, we ignore reflections from the substrate.  To maximize the photoelastic modulation, we performed a polarization sweep, varying the input laser linear polarization direction in the X-Y plane. The measured signal power exhibited a $90^\circ$ periodicity, with a maximum at polarization along the crystal X-axis (utilizing the photoelastic coefficient $p_{12}$) and a minimum along Y-axis (utilizing $p_{11}$), showing a 4.7~dB difference. This matches the ratio $(p_{12}/p_{11})^2$, as expected from Eq.~\ref{refractive_index}. Therefore, the input laser was chosen to be polarized along the quartz crystal X-axis for the data presented in the main text.  We approximate the input optical signal as a plane wave and write the reflected optical waves from the front surface $\vec{E}_1$ and back surface $\vec{E}_2$ as:
\begin{align}
    \vec{E}_1=c_1e^{i(kz-\omega t)}\hat{x},\ 
    \vec{E}_2=c_2e^{i(kz-\omega t-\delta)}\hat{x}    
\end{align}
where $\delta$ is their phase difference:
\begin{align}\label{phase_difference}
    \delta(x,y,t) = n_x(x,y,t)\,\frac{\omega}{c}\cdot 2T=\delta_0-M\,\sin(\omega_m t)
\end{align}
with $T$ being the thickness of the resonator along the input laser path and
\begin{align}
    \delta_0&\equiv n_o\cdot \frac{\omega}{c}\,2T,
    \ M\equiv\frac{\omega}{c}T\cdot n_o^3\,p_{12}\cdot S_0\,\cos(\frac{\pi y}{L})
\end{align}
Rigorously, the phase difference should be $2\int n_x\,\frac{\omega}{c}\,dz$ integrated along the optical path, but for our mode, the strain $S_{yy}$ (hence $n_x$) is relatively uniform along the Z-axis.

The optical power detected by the photodetector is:
\begin{align}\label{eqn:C13}
    I&\propto\frac{||\Vec{E}_1+\Vec{E}_2||^2}{2}=\frac{1}{2}\left(c_1^2+c_2^2+2c_1c_2\cos\delta\right)\notag\\
    &\approx \frac{1}{2}\,[c_1^2+c_2^2+2\,c_1c_2\cos\delta_0\,J_0(M)\notag\\
    &\quad\ +4\,c_1c_2\sin\delta_0\,J_1(M)\,\sin(\omega_mt)]
\end{align}
where the approximation for the last step is made assuming that $M$ is small and $J_\alpha(x)$ is the Bessel function of the first kind.  Since $J_1(M)\approx \frac{M}{2}\propto u_0$ (for small $M$), the depth of modulation is proportional to the amplitude of the mechanical motion. Hence, the mechanical response is linearly transduced into our optical response.

\begin{figure}[h]
    \centering
    \includegraphics[width=\linewidth]{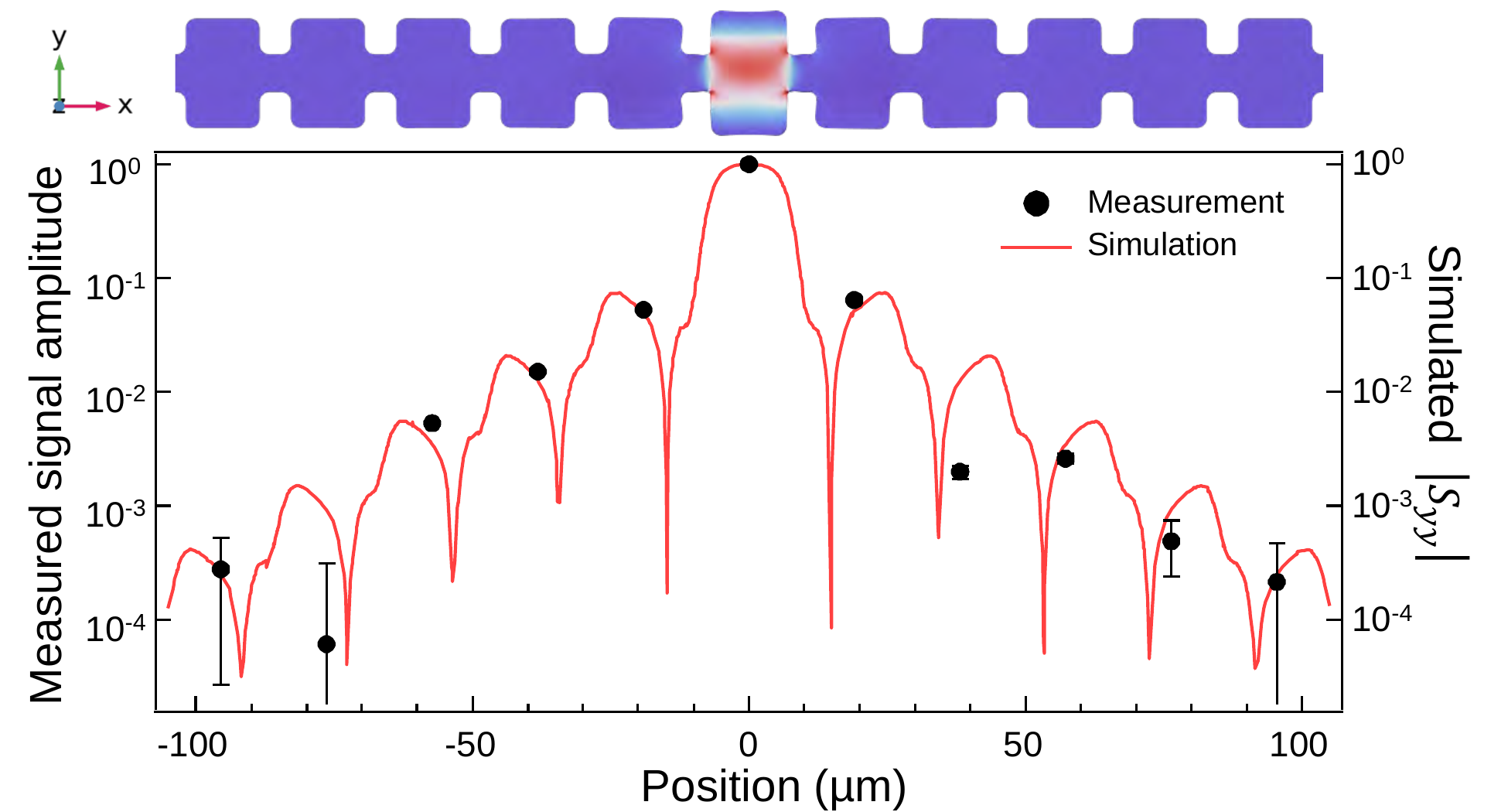}
    \caption{\textbf{Mechanical width-extension mode profile.} Upper panel shows the simulated $S_{yy}$ strain distribution of a X chain (on the left is the crystal axes orientation). Lower panel shows the maximum signal amplitude at each unit cell normalized to 1 at the central defect (Black dots), and the normalized simulated strain distribution ($|S_{yy}|$) along the central axis line of our 5M resonator (Red curve).}
    \label{Signal_spatial_distribution}
\end{figure}

Utilizing our platform for photoelastic imaging, allows us to profile the mechanical mode shape.  We scan the laser spot position while keeping the incident optical power constant. Here, the microwave probe drives the mechanical mode resonantly with a fixed position and exciting power. Figure~\ref{Signal_spatial_distribution} shows the measured profile of our width-extension mode and the simulated strain, $S_{yy}$. The measured signal strength shows a rapid exponential decay into the phononic crystal shielding, indicating a strongly localized defect mode.

\begin{figure}[h]
    \centering
    \includegraphics[width=0.9\linewidth]{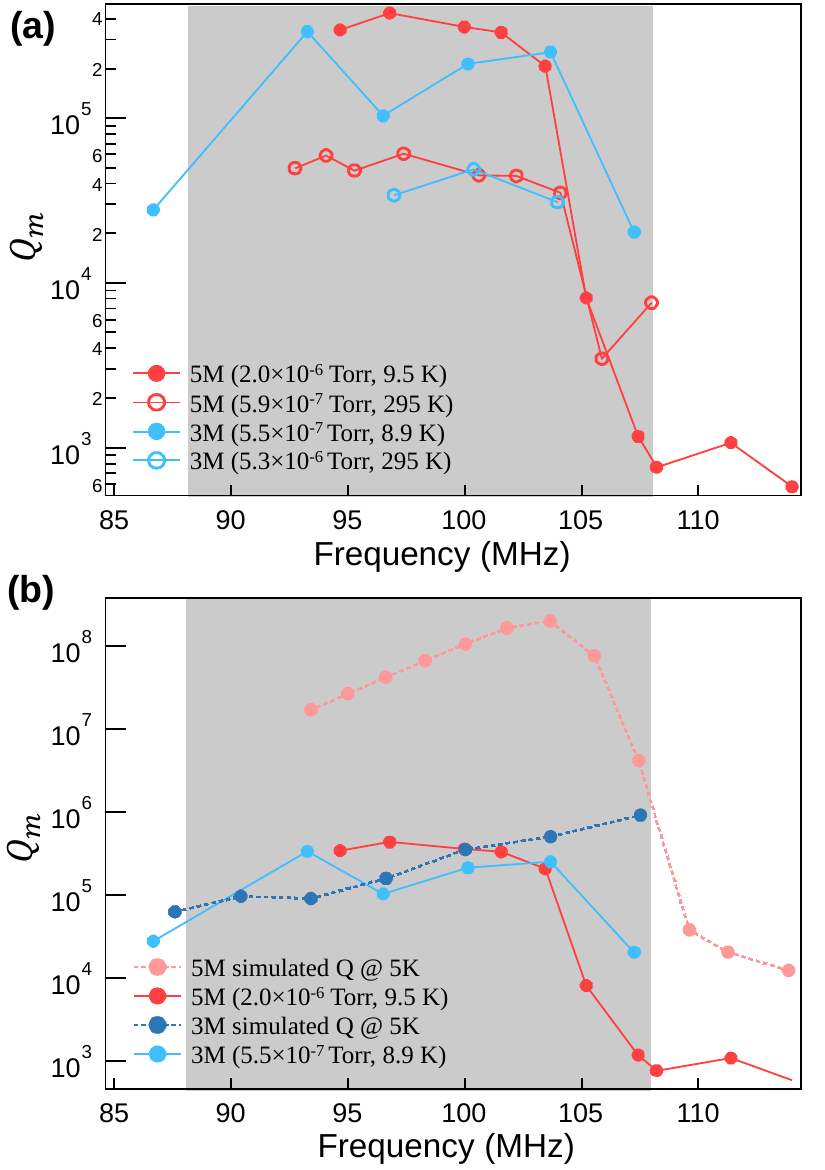}
    \caption{\textbf{Comparison of 3M and 5M resonators.} (a) Measured Q of 3M and 5M resonators at room temperature and lowest reached cryogenic temperature. (b) Measured Q at cryogenic temperature and simulated radiation Q of 3M and 5M resonators.}
    \label{3M vs 5M}
\end{figure}

\section{3M array vs 5M array}\label{Appendix:3M_array}

Figure~\ref{3M vs 5M} compares the performance of resonators in the 3M and 5M resonator arrays, with the gray-shaded area representing the simulated target acoustic band gap. Figure~\ref{3M vs 5M}(a) shows the measured Q at room temperature and cryogenic temperature for both 3M and 5M resonators. Both exhibit similar maximum Q at both room temperature and cryogenic temperatures. Figure~\ref{3M vs 5M}(b) compares the Q measured at cryogenic temperature with the simulated radiation Q as a function of mode frequency. While simulations predict an increase in radiation Q as the mode frequency enters deeper into the acoustic band gap, the measured Q for the 5M resonators instead plateaus at a value significantly lower than the simulated limit. This indicates that anchor losses are not limiting our Q. The 3M resonators are of a slightly different thickness from the 5M resonators, due to etching non-uniformity. The difference in thickness from the target value alters the width and center of the band gap, somewhat lowering the efficacy of the acoustic shielding.

\begin{figure}[h]
    \centering
    \includegraphics[width = \linewidth]{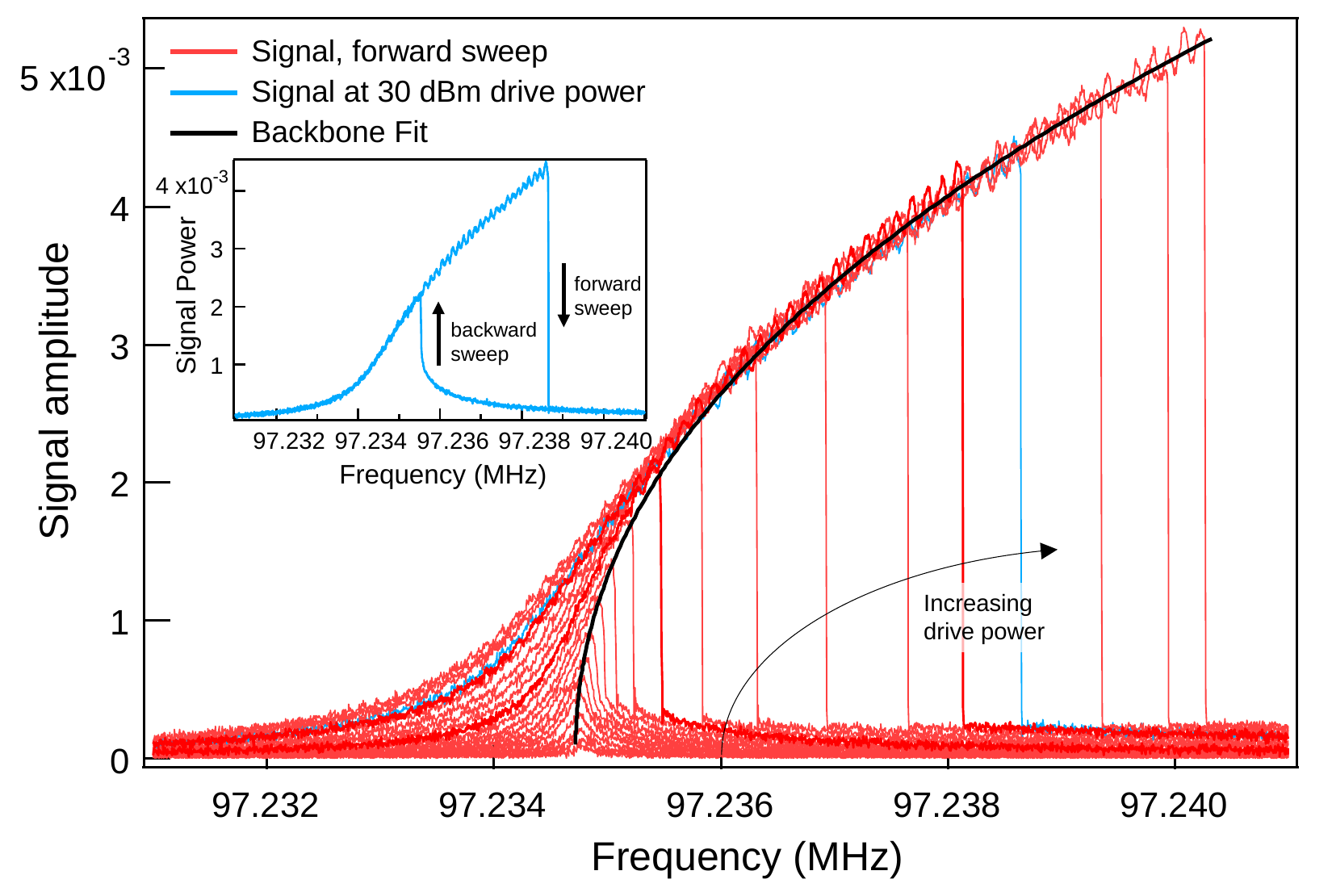}
    \caption{\label{Nonlinearity}\textbf{Mechanical nonlinearity.} Mechanical signal measured by the VNA during forward frequency sweeps at various driving powers from -10~dBm to 33~dBm. The black backbone curve traces the peaks of each response curve. To illustrate the hysteretic behavior in the nonlinear region, both forward and backward frequency sweep curves for 30~dBm driving power are shown in the inset.}
\end{figure}

\section{Mechanical nonlinearity}\label{Appendix:Mechanical_Nonlinearity}

At cryogenic temperatures where the resonator's Q is high, mechanical nonlinearity becomes prominent at large excitation power, and the mechanical response is no longer Lorentzian as shown in Fig.~\ref{Nonlinearity}. Such phenomenon is well characterized by a Duffing-type nonlinearity \cite{Nonlinearity_Lifshitz_Cross_2010}.  Analysis of a driven Duffing resonator exhibits a bistable region of the mechanical response that results in hysteretic behavior as shown in the inset of Fig.~\ref{Nonlinearity}. The driving power stiffens our mechanical resonator, yielding a higher mode frequency. The peaks of curves with different driving power form a backbone curve, which can be fitted to the equation \cite{kozinsky_tuning_2006}:
\begin{equation}
    f = f_0 + A\,a^n
\end{equation}
with $f_0=97.2~\text{MHz},A=5.12\times10^8~\text{Hz},n=2.17$ with $a$ being the magnitude of the signal measured by the photodetector, which is, from Eq.~\ref{eqn:C13}, proportional to the amplitude of mechanical motion.  This behavior is well outside the single-phonon regime of the operation of our device as a quantum memory.

\section{Surface adsorption}\label{Appendix:Surface_adsorption}

\begin{figure}[h]
    \centering
    \includegraphics[width=\linewidth]{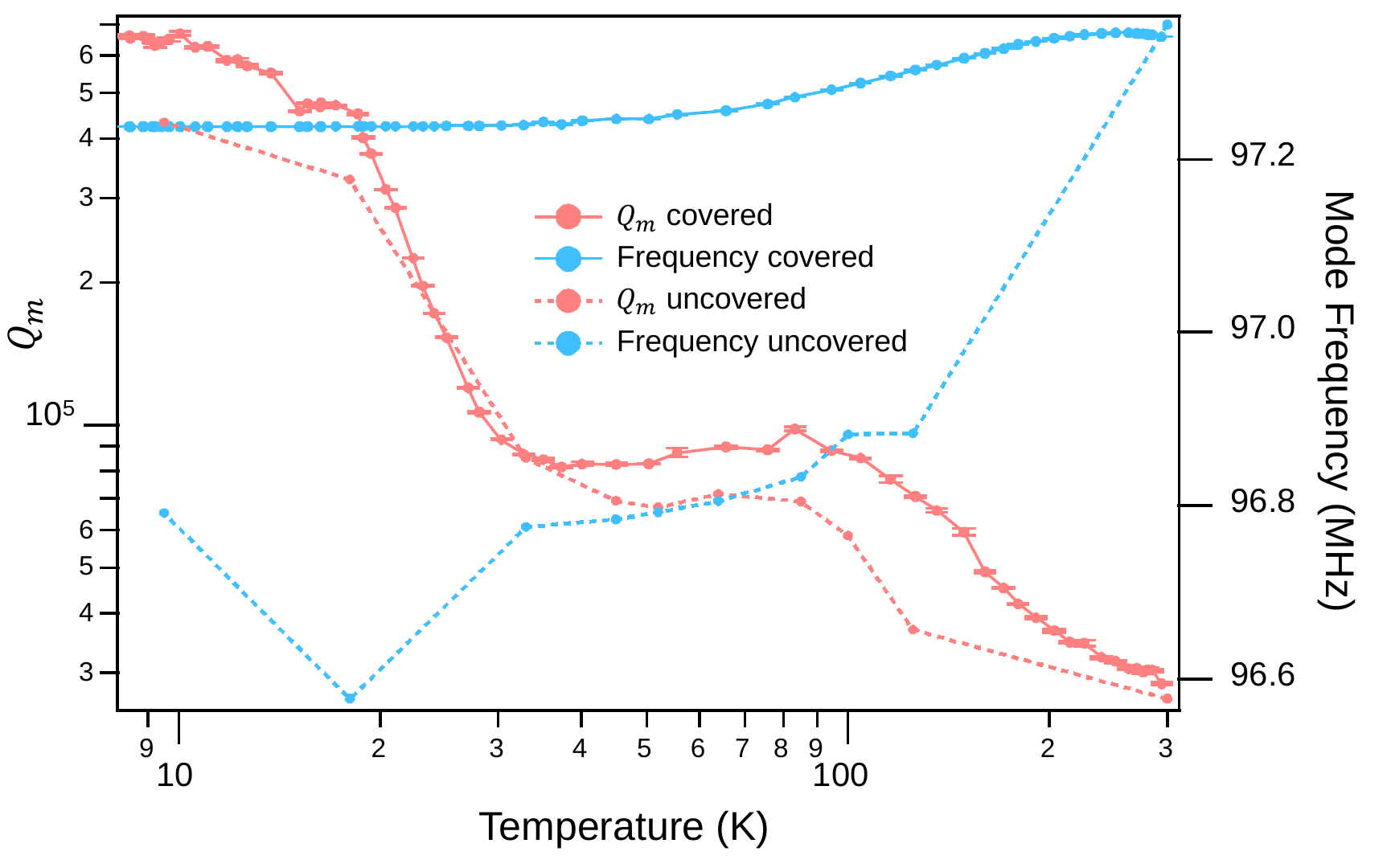}
    \caption{\textbf{Effect of particle adsorption on mechanical Q and mode frequency.} Quality factor and mode frequency versus temperature for the max-Q mode for experimental cooldowns with and without the cover chip are shown.}
    \label{Covered_vs_Uncovered}
\end{figure}

\begin{figure}[h]
    \centering
    \includegraphics[width=\linewidth]{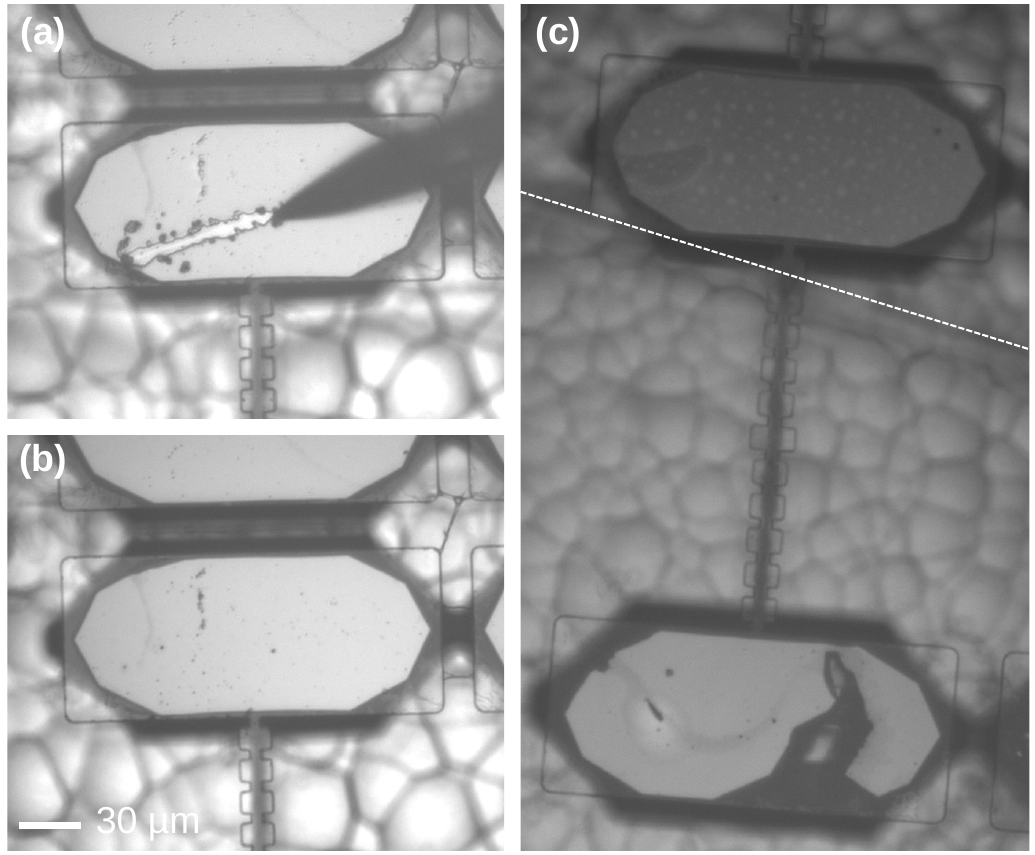}
    \caption{\textbf{Surface adsorption.} (a) Visible scratch after dragging the electric probe across the resonator at cryogenic temperature. (b) After warming up, the scratch in (a) disappears. (c) Resonator with quartz cover chip about 70~$\mu$m above. White dashed line indicates the edge of the cover chip. Picture is taken at around 263~K during warm up, where liquid droplets are clearly seen on the uncovered part of the resonator.}
    \label{Particle adsorption}
\end{figure}

During our initial cooldown, we observed a substantial redshift in the mechanical mode frequencies.  However, the temperature dependence of the material properties of bulk quartz suggests that our devices should slightly stiffen and increase their frequencies at cryogenic temperatures.  Suspecting that a mass loading effect from residual gas frozen on the surface of our device caused such frequency shifts, we covered our resonators with a thin quartz chip to shield them from contamination (and blackbody radiation) during subsequent cooldowns (detailed in the Experimental Setup section).

Figure~\ref{Covered_vs_Uncovered} shows the Q and the mode frequency as function of temperature with and without the cover chip. With the cover chip, the maximum Q increased and the mode frequency redshift was significantly suppressed. During our initial cooldown, we were also able to directly image our resonators through the optical window of the cryostat. After cooling to the lowest temperature, we noticed that dragging the electric probe across the surface of the die left visible marks on the surface (Fig.~\ref{Particle adsorption}(a)). During the warm-up process, these scratch marks disappeared and visible liquid droplets nucleated at $\sim$230~K which completely evaporated by around $\sim$275~K (Fig.~\ref{Particle adsorption}(b)), suggesting water and other contaminant adsorption. Figure~\ref{Particle adsorption}(c) shows a partially covered resonator during warm up from cryogenic temperature, where the uncovered upper region (upper anchor pad) exhibits liquid droplets while the covered lower region remains clean.  Droplets on the top surface of the cover chip are out of focus and not visible in the image.

\bibliography{apssamp}
\end{document}